\documentclass[aps,prd,10pt,twocolumn,superscriptaddress,showpacs,footinbib]{revtex4-1}
\pdfoutput=1
\usepackage{hyperref}
\usepackage{slashed}
\usepackage{bm}
\usepackage{graphicx}
\usepackage{gensymb}
\usepackage{enumerate}
\usepackage{amsfonts,amsmath,amssymb,bm,bbm}
\usepackage{color}
\usepackage{footnote}
\usepackage{epsfig, subfigure}
\usepackage{setspace}
\usepackage{booktabs, tabularx} 
\usepackage{multirow}

\hypersetup{
    pdfnewwindow=true,      
    colorlinks=true,       
    linkcolor=blue,          
    citecolor=blue,        
    filecolor=blue,      
    urlcolor=blue        
}

\newcommand{\be}{\begin{equation}}  
\newcommand{\ee}{\end{equation}}
\newcommand{\bea}{\begin{eqnarray}}  
\newcommand{\eea}{\end{eqnarray}}

\begin{document}

\title{The effect of hydrodynamical simulation inspired dark matter velocity profile on directional detection of dark matter}

\author{Ranjan Laha}
\affiliation{Kavli Institute for Particle Astrophysics and Cosmology (KIPAC),
	\\ Department of Physics, Stanford University, Stanford, CA 94305, USA
 \\ SLAC National Accelerator Laboratory, Menlo Park, CA 94025, USA \\
{\tt ranjalah@uni-mainz.de}\smallskip} 

\date{\today}
\begin{abstract}
Directional detection is an important way to detect dark matter.  An input to these experiments is the dark matter velocity distribution.  Recent hydrodynamical simulations have shown that the dark matter velocity distribution differs substantially from the Standard Halo Model.  We study the impact of some of these updated velocity distribution in dark matter directional detection experiments. We calculate the ratio of events required to confirm the forward-backward asymmetry and the existence of the ring of maximum recoil rate using different dark matter velocity distributions for $^{19}$F and Xe targets.  We show that with the use of updated dark matter velocity profiles, the forward-backward asymmetry and the ring of maximum recoil rate can be confirmed using a factor of $\sim$2 -- 3 less events when compared to that using the Standard Halo Model.
\end{abstract}

\maketitle

\section{Introduction}
\label{sec:introduction}

Despite the overwhelming astrophysical evidence for dark matter, particle physics signatures of dark matter are still lacking\,\cite{Planck:2015xua,Steigman:2007xt,Strigari:2014yea,Bhattacharjee:2012xm}.  There are various ways to detect dark matter particle candidates with masses GeV $\lesssim \, m_\chi \lesssim$ TeV.  Direct detection, indirect detection, and collider searches form the three-prong approach to detect dark matter particles in this mass range\,\cite{Undagoitia:2015gya,Boveia:2016mrp,Klasen:2015uma}.  

Among these three search strategies, direct detection of dark matter is the only way to detect local dark matter particles\,\cite{Catena:2011kv}.  These searches typically proceed via the detection of $\sim \mathcal{O}$(keV) nuclear recoils.  Due to the enormous background at these energies, it is extremely difficult to distinguish the dark matter signal from background.  In past, dark matter signals have been claimed by some of these searches, however, none of these have stood further detailed scrutiny\,\cite{Herrero-Garcia:2015kga,DelNobile:2015lxa,Catena:2016hoj,Scopel:2015eoh,Yang:2016wrl}.  

In order to separate signal from background, it was pointed out some time ago to utilize the directional nature of the scattering of dark matter particle with nuclei\,\cite{Spergel:1987kx}.  The motion of the Solar system through the Galaxy will produce a distinct angular recoil spectrum\,\cite{Copi:1999pw,Copi:2000tv,Gondolo:2002np,Morgan:2004ys,Billard:2009mf,Ahlen:2009ev,Green:2010zm,Lee:2012pf,Grothaus:2014hja,O'Hare:2014oxa,O'Hare:2015mda,Laha:2015yoa,Mayet:2016zxu,Kavanagh:2015aqa,Kavanagh:2016xfi}.  It is expected that background will not produce such an angular recoil spectrum.  

There are numeorus ongoing directional dark matter detection experiments, for e.g., DRIFT\,\cite{Daw:2011wq,Battat:2014van}, D3\,\cite{Jaegle:2011rn,Vahsen:2011qx}, DMTPC\,\cite{Monroe:2012qma,Monroe:2011er,Leyton:2016nit}, NEWAGE\,\cite{Miuchi:2010hn,Miuchi:2011qw}, and MIMAC\,\cite{Riffard:2013psa,Riffard:2016mgw}. All these experiments need to reconstruct a track of length $\sim$ $\mathcal{O}$(mm).  All these gaseous targets have a small target mass, and scaling up to a sizable target mass is also an enormous challenge\,\cite{Naka:2013nla,Cappella:2013rua,Capparelli:2014lua,Aleksandrov:2016fyr}.  Recently, there have been suggestions to use dense Xenon gas as a target for directional dark matter detection, but the research and development in that direction is still in a very nascent stage\,\cite{Nygren:2013nda,Gehman:2013mra,Mohlabeng:2015efa,Li:2015zga}.  

In addition to the forward-backward asymmetry, a ring like feature can also be used as an efficient discriminator between signal and background in a dark matter directional detection experiment\,\cite{Bozorgnia:2011vc}.  The ring corresponds to the angle at which the angular recoil rate is the maximum.  The angular recoil rate has a maximum at the ``ring angle" and falls off at angles away from it, and this maximum rate appears as a ring (due to the azimuthal symmetry of the scattering) when viewed in 3 dimensions.  This feature prominently appears for a dark matter particle masses $\gtrsim$ 100 GeV, and for a low nuclear recoil threshold\,\cite{Bozorgnia:2011vc}.  Due to the importance of this feature, it is imperative to check the robustness of this feature for various different dark matter velocity distributions.  The ring feature is also present for bound state dark matter ``darkonium" and the conclusions in this work apply qualitatively for it too\,\cite{Laha:2015yoa,Laha:2013gva}.

In this work, we investigate the forward-backward asymmetry and the ring for updated dark matter velocity profiles.  Recent hydrodynamical dark matter simulations have shown that the Milky Way dark matter velocity profile deviates from the Standard Halo Model (SHM)\,\cite{Ling:2009eh,Kuhlen:2013tra,Butsky:2015pya,Bozorgnia:2016ogo}.  The impact of these velocity profiles on non-directional dark matter searches have been considered recently\,\cite{Bozorgnia:2016ogo,Kelso:2016qqj,Sloane:2016kyi,Petersen:2016vck}.  Current constraints on dark matter - nucleon scattering cross section from directional detection experiments are quite weak.  We estimate the ratios of the number of events required to reach 3$\sigma$ discrimination in forward-backward asymmetry and the appearance of a ring for various different velocity profiles.  Using the ratio makes our result independent of the uncertainties due to the dark matter local density, and the dark matter - nucleon cross section.  We remind the reader that in this work, we are only explore the magnitude of the dark matter velocity, i.e., the speed distribution.  We use the word ``velocity" following convention.

We show our results for two targets: $^{19}$F and Xe.  Although other targets are also used in directional detection, our choice is representative, and bracket the uncertainty due to different nuclear targets.  

The remaining part of the work is arranged as follows.  In Section\,\ref{sec:calculation}, we introduce the various dark matter velocity profiles, and recapitulate the necessary formulas for dark matter directional detection.  We present our results in Sec.\,\ref{sec:results}, and conclude in Sec.\,\ref{sec:conclusion}.

\section{Calculations}
\label{sec:calculation}

\subsection{Dark matter velocity profile}
\label{sec:dark matter velocity profile}

The typical dark matter velocity profile used is the standard halo model:
\begin{eqnarray}
f(v) \propto \dfrac{1}{(2\pi \sigma_v^2)^{3/2}} \, e^{- v^2/2 \sigma_v^2} \, ,
\label{eq:SHM}
\end{eqnarray}
where $\sigma_v$ = 155.59 km s$^{-1}$.  This analytical velocity model arises from the assumption of an isothermal dark matter density profile.  The dark matter velocity in the inertial Galactocentric frame is denoted by $v$.  The escape velocity is assumed to be $v_{\rm esc} \sim 600$ km s$^{-1}$\,\cite{Piffl:2013mla}.  Hydrodynamical simulations which include baryons give a different dark matter velocity profile.  Recently, Milky Way like halos from the EAGLE HR\,\cite{Schaye:2014tpa,Crain:2015poa} and APOSTLE IR\,\cite{Sawala:2015cdf,2016MNRAS.457..844F} simulations were fit to four different dark matter velocity profiles:
\newline{(}1) standard Maxwellian distribution:
\begin{eqnarray}
f (v) \propto v^2 \, {\rm exp}\,[-(v/v_0)^{2}] \, ,
\label{eq:standard Maxwellian distribution}
\end{eqnarray}
\newline{(}2) generalized Maxwellian distribution:
\begin{eqnarray}
f (v) \propto v^2 \, {\rm exp}\,[-(v/v_0)^{2\alpha}] \, ,
\label{eq:generalized Maxwellian distribution}
\end{eqnarray}
\newline{(}3) velocity distribution advocated by Mao et al.\,\cite{Mao:2012hf}:
\begin{eqnarray}
f (v) \propto v^2 \, {\rm exp}\,[-v/v_0] \, (v_{\rm esc}^2 - v^2)^p \, \Theta (v_{\rm esc} - v) \, ,
\label{eq:Mao etal velocity distribution}
\end{eqnarray}
\newline{a}nd (4) velocity distribution advocated by Lisanti et al.\,\cite{Lisanti:2010qx}:
\begin{eqnarray}
f (v) \propto v^2 \, {\rm exp}\,[(v_{\rm esc}^2 - v^2)/(k v_0^2 - 1)]^k \, \Theta (v_{\rm esc} - v) \, .
\label{eq:Lisanti etal velocity distribution}
\end{eqnarray}


\begin{figure}
\includegraphics[angle=0.0,width=0.45\textwidth]{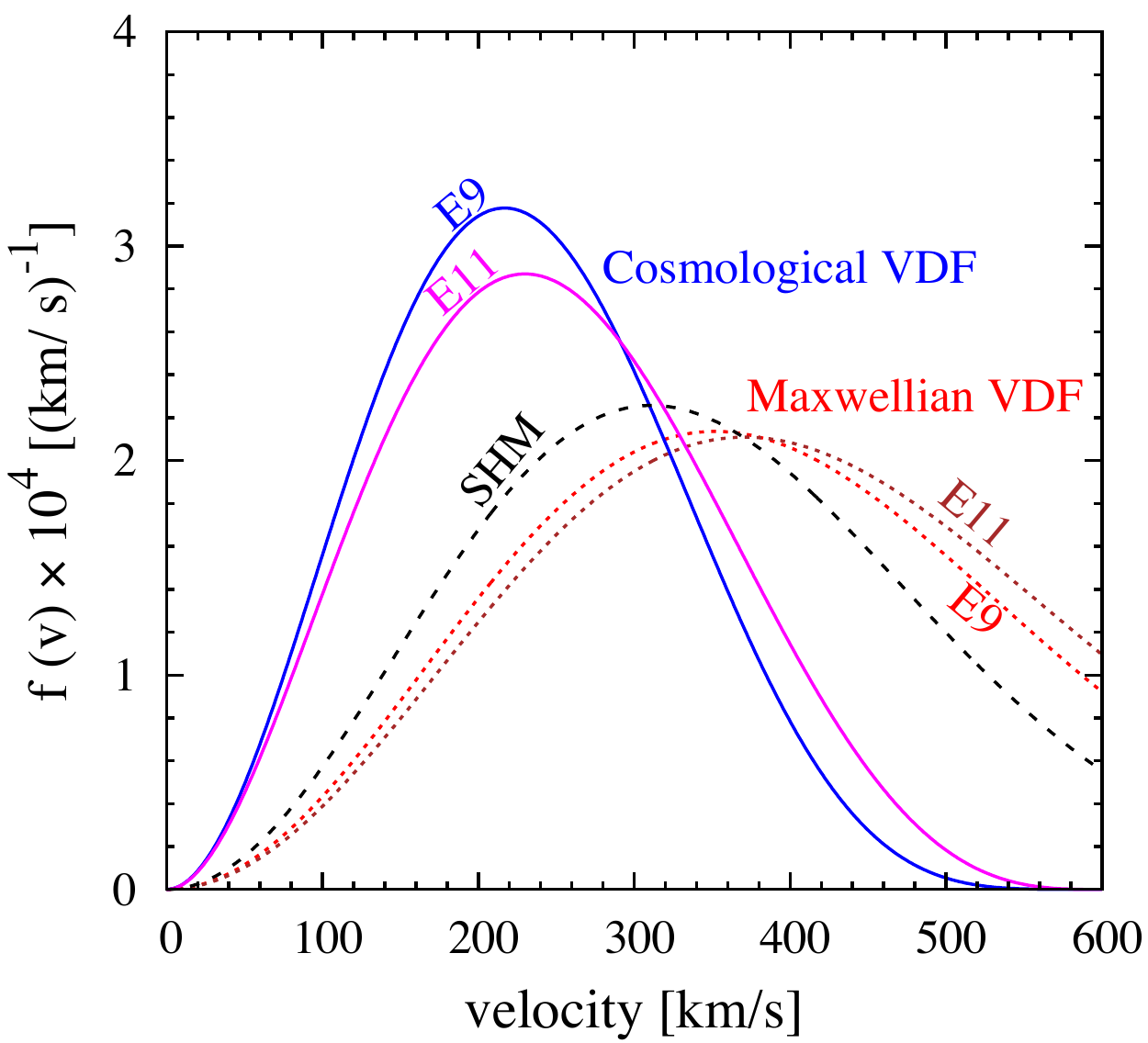}
\caption{The Milky Way dark matter velocity profiles considered in this work.  The Standard Halo Model is labelled as SHM.  The standard Maxwellian velocity distribution (eqn.\,\ref{eq:standard Maxwellian distribution}) fits to halos E9 and E11 are labelled as Maxwellian VDF.  The Mao et al. velocity distribution (eqn.\,\ref{eq:Mao etal velocity distribution}) fits to halos E9 and E11 are labelled as Cosmological VDF.} 
\label{fig:Comparison velocity profile}
\end{figure}


The criteria for the selection of Milky Way like halos from the simulations were: $(i)$ agreement with the observed Milky Way rotation curve, $(ii)$ stellar mass similar to the Milky Way: 4.5 $\times$ 10$^{10}$ $M_\odot$ $<$ $M_*$ $<$ 8.3 $\times$ 10$^{10}$ $M_\odot$, and $(iii)$ the presence of a stellar disc\,\cite{Calore:2015oya}.  There were 14 halos which fit the first two criteria, and only two halos, E9 and E11, fit all the criteria.  In general, all of these halos are better fit by the Mao et al. velocity profile.  We concentrate on the standard Maxwellian and the Mao et al. velocity profiles as derived for the two halos: E9 and E11.  The parameters are:

(i) E9: standard Maxwellian $v_0 = 248.81$ km s$^{-1}$; \,\,\,Mao et al. $v_0 = 393.63$ km s$^{-1}$, and $p$ = 4.82, 

$(ii)$ E11: standard Maxwellian $v_0 = 262.27$ km s$^{-1}$; Mao et al. $v_0 = 250.06$ km s$^{-1}$, and $p$ = 3.14. 

These velocity profiles are shown in Fig.\,\ref{fig:Comparison velocity profile}.  We denote the standard Maxwellian velocity distribution function as Maxwellian VDF, and the Mao et al. velocity distribution function as Cosmological VDF.  For comparison, the standard halo model is also shown as SHM.  The Maxwellian VDF deviates substantially from the Cosmological VDF for both the halos E9 and E11.  This is a reflection of the poor reduced $\chi^2$ for the Maxwellian VDF for both these halos.  Inspite of the poor fit, we include the Maxwellian VDF to broadly encompass the uncertainties in the dark matter velocity profile.    

\subsection{Dark matter directional detection}
\label{sec:dark matter directional detection}

The formalism for dark matter directional detection is well known.  Here we recapitulate the main ideas for completeness.  The double differential rate ($R$) w.r.t. the nuclear recoil energy ($E_{\rm nr}$) and solid angle $(\Omega)$ of a dark matter particle colliding with a nucleus is given by\,\cite{Laha:2015yoa} 
\begin{eqnarray}
\dfrac{d^2 R}{dE_{\rm nr} \, d\Omega} &&= N_T \, n_\chi \int d^3 {\bf v} \, f(v) \nonumber\\
&\times& \dfrac{\sigma^{\rm SD}_A \, F^2_{\rm SD} (E_{\rm nr}) \, m_A}{4\pi \, \mu^2} \,\delta \left({\bf v}.\hat{\bf{q}} - \dfrac{q }{2 \mu }\right) \, ,\phantom{111}
\label{eq:dark matter elastic Galactic frame}
\end{eqnarray}  
where $N_T$ denotes the number of target nuclei, the local number density of dark matter particles is denoted by $n_\chi$, $\sigma_A^{\rm SD}$ denotes the spin-dependent dark matter - nucleon cross section, $F_{\rm SD}^2$ denotes the spin-dependent nuclear form factor, $m_A$ denotes the mass of the target nuclei, $\mu$ denotes the reduced mass of the dark matter - nucleus system, $\bf{v}$ is the dark matter velocity vector in the Galactic frame, $\bf{q}$ denotes the nuclear recoil direction vector with $\hat{\bf{q}}$ being the corresponding unit vector.  We have chosen the spin-dependent cross section in this expression as traditionally directional detection experiments show constraints for this interaction.  This choice has little effect on the main results presented in the paper.  Recent theoretical work has also considered the effect of dark matter effective operators on the various directional features in a dark matter experiment\,\cite{Catena:2015vpa,Kavanagh:2015jma}.  

Transforming this expression to the laboratory frame gives us
\begin{eqnarray}
\dfrac{d^2 R}{dE_{\rm nr} \, d\Omega_{v_E q}} &&= N_T \, n_\chi \int _{v_E \, {\rm cos} \, \theta_{v_Eq} + q/2\mu} ^{v_{\rm max}} \dfrac{\sigma^{\rm SD}_A \, F^2_{\rm SD} (E_{\rm nr}) \, m_A}{4\pi \, \mu^2} \nonumber\\
&\times& 2\pi \,v \, f(v) \, dv \, ,
\label{eq:eq:dark matter elastic Galactic frame integrated over theta_vq}
\end{eqnarray}
where $\Omega_{v_Eq}$ denotes that the solid angle is between the velocity of the Earth and the nuclear recoil direction, and $v_{\rm max}$ denotes the maximum velocity of the dark matter particles.  When the dark matter velocity distribution follows the standard Maxwellian velocity distribution, the above mentioned equation can be integrated exactly to obtain
\begin{eqnarray}
&&\dfrac{d^2 R}{dE_{\rm nr} \, d\Omega_{v_E q}} = N_T \, n_\chi  \dfrac{\sigma^{\rm SD}_A \, F^2_{\rm SD} (E_{\rm nr}) \, m_A}{4\pi \, \mu^2} \nonumber\\
&\times&  2\pi N\, v_0^2 \left(e^{-\dfrac{(v_E \, {\rm cos} \, \theta_{v_Eq} + q/2\mu)^2}{2v_0^2}} - e^{-\dfrac{v_{\rm max}^2}{2 v_0^2}} \right),
\label{eq:dark matter elastic Galactic frame final expression}
\end{eqnarray}
where the speed of the Earth w.r.t. the Galaxy is denoted by $v_E$.  The normalization constant for the velocity distribution is denoted by $N = 1/4 \pi \,\times 1/(N_1 + N_2)$, where
\begin{eqnarray}
&N_1& \, =\,  - v_{\rm max} \, v_0^2 \, {\rm exp}\left(-\dfrac{v_{\rm max}^2}{2 \, v_0^2} \right) \,, \\
&N_2& \, =\, \sqrt{\dfrac{\pi}{2}} \, v_0^3 \, {\rm erf} \left(\dfrac{v_{\rm max}}{\sqrt{2} v_0} \right) \,,
\label{eq:normalization constants}
\end{eqnarray}
where erf denotes the error function.

\begin{figure}
\includegraphics[angle=0.0,width=0.99\columnwidth]{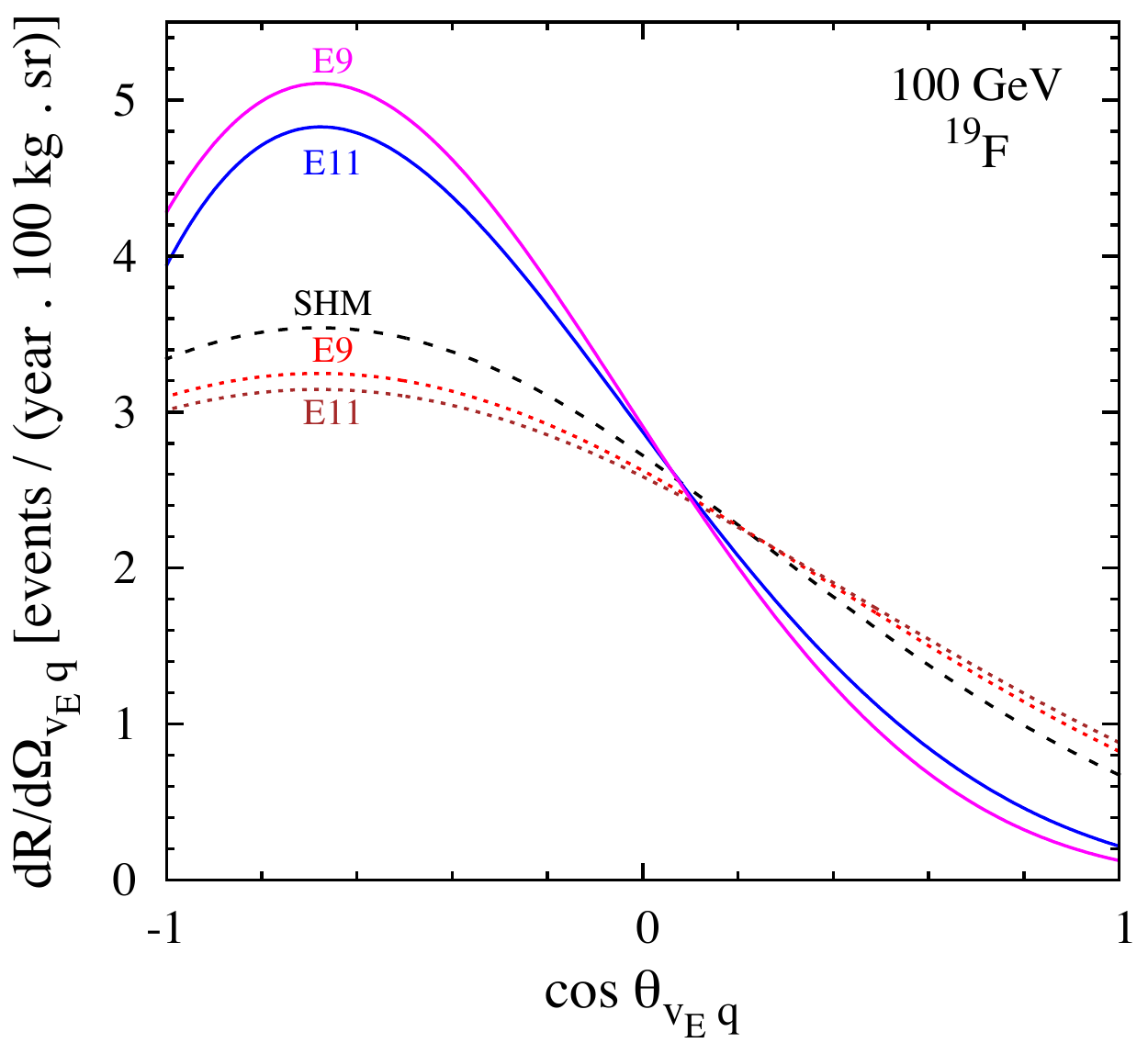}
\includegraphics[angle=0.0,width=0.99\columnwidth]{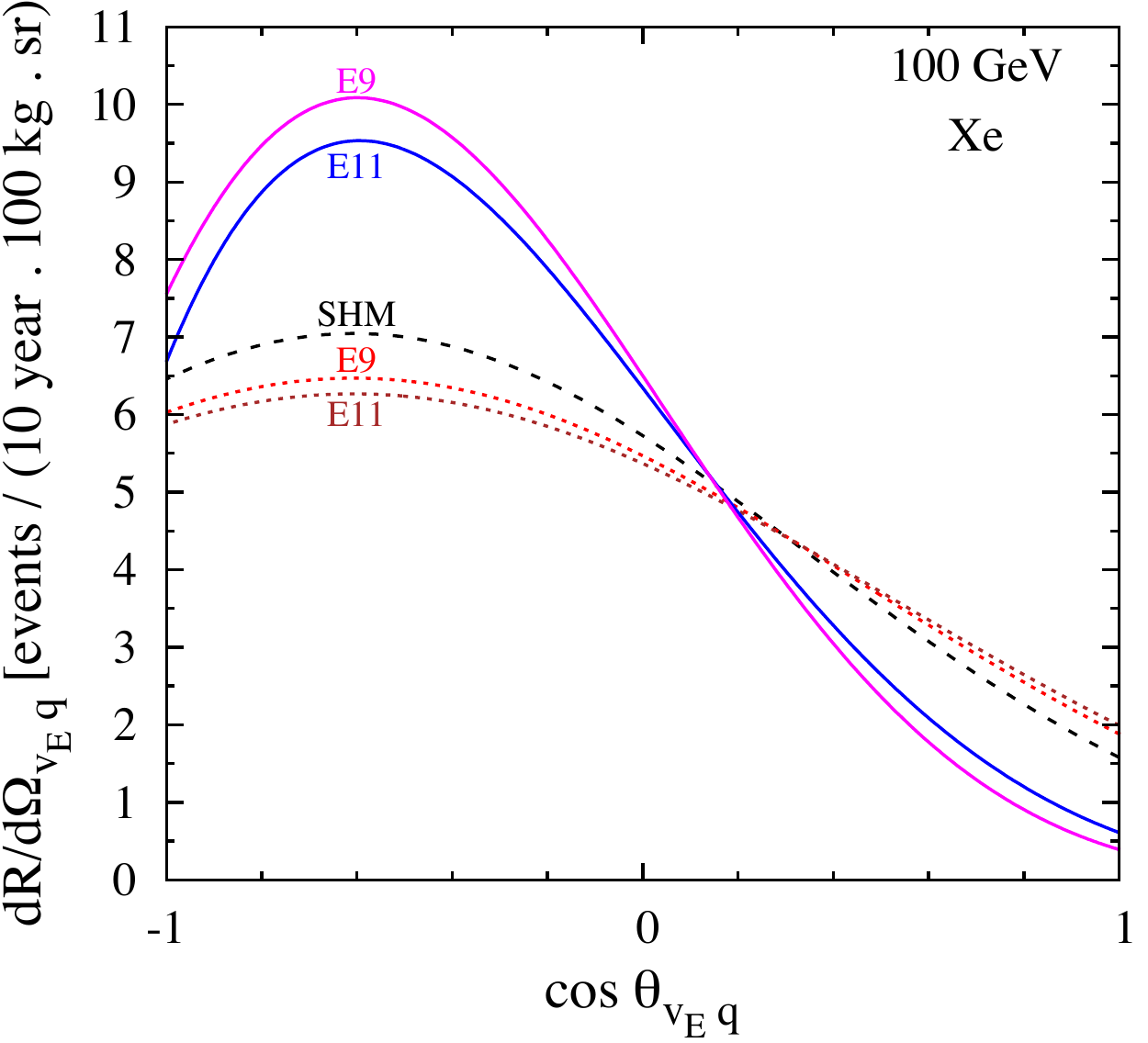}
\caption{The differential angular nuclear recoil spectrum for different Milky Way dark matter velocity distribution.  The velocity profiles considered are the Mao et al. profile (solid blue and solid magenta) fit to halos E9 and E11, the SHM velocity distribution (black dotted), and the standard Maxwellian fit (red dotted and brown dotted) fit to halos E9 and E11.  The target in the top panel is $^{19}$F, and that in the bottom panel is Xe.  We consider spin-dependent interactions for both this plots.  The integrated energy range considered for $^{19}$F and Xe is [5, 10] keV, and [5, 20] keV respectively.}
\label{fig:differential recoil rate}
\end{figure}


Both the forward-backward asymmetry and the ring-like structure can be understood as a competition between the two exponential functions in eqn.\,\ref{eq:dark matter elastic Galactic frame final expression}.  For the dark matter masses that we consider and the $^{19}$F nuclei, the values of $\mu$ varies from $\sim$16 GeV to $\sim$18.6 GeV.  For the dark matter masses that we consider and the Xe nuclei, the values of $\mu$ varies from $\sim$56.3 GeV to $\sim$115.8 GeV.  For the recoil energies that we consider, the recoil momentum of the $^{19}$F nuclei falls between $\sim$13.7 MeV and $\sim$19.5 MeV.  The recoil momentum of the Xe nuclei falls between $\sim$32 MeV and $\sim$72.4 MeV.  

The various features in a directional detection experiment can be understood by analyzing eqn.\,\ref{eq:dark matter elastic Galactic frame final expression}.  The dependence on the angle, $\theta_{v_Eq}$, arises through the first exponential term.  The energy dependence arises through the spin-dependent form factor and the first exponential term.  However, the energy dependence of the spin-dependent form factor is weak, especially for the low nuclear recoil energy, and it can be approximated as 1.

The number of events in the forward region, ${\rm cos}\,\theta_{v_Eq} \leq 0$, denoted by $N_F$ is larger than that in the backward region, ${\rm cos}\,\theta_{v_Eq} \geq 0$ denoted by $N_B$.  This simply follows from eqn.\,\ref{eq:dark matter elastic Galactic frame final expression} where one can show that at a given energy and angle
\begin{eqnarray}
\dfrac{\rm Forward \, differential \, rate}{\rm Backward \, differential \, rate} \approx \dfrac{e^{-\dfrac{(- v_E \, |{\rm cos} \, \theta_{v_Eq}| + q/2\mu)^2}{2\,v_0^2}}}{e^{-\dfrac{(v_E \, |{\rm cos} \, \theta_{v_Eq}| + q/2\mu)^2}{2\,v_0^2}}} \, . \nonumber\\
\label{eq:forward-backward ratio}
\end{eqnarray}
In the above approximate expression, we have neglected the term exp $[-v_{\rm max}^2/(2 \, v_0^2)] \ll$ 1.

The position of the maximum of the angular recoil rate can be derived from the term exp$\left[-\dfrac{(v_E \, {\rm cos}\,\theta_{v_Eq} + q/2\mu)^2}{2\,v_0^2}\right]$.  If $q/2\mu < v_E$, the maximum of the exponential happens when the numerator in the argument becomes zero.  This happens at cos\,$\theta_{v_Eq} = -q/(2\mu \, v_E)$ which gives the ``ring angle".  The condition for the ``ring angle" also implies that the ring is visible for heavier dark matter masses and lower recoil energies.  The contrast between the differential rate at the ring, and that at the cos\,$\theta_{v_Eq}$ = 1 can be analytically derived as
\begin{eqnarray}
{\rm Ring \, constrast} \approx \dfrac{1}{e^{-\dfrac{(v_E + q/2\mu)^2}{2\, v_0^2}}} \, .
\label{eq:ring contrast}
\end{eqnarray}
This shows that dark matter velocity profiles with smaller $v_0$ can produce a larger contrast in the ring.

An experimental detection of the forward backward asymmetry and the ring depends on both the sense recognition and the angular resolution of the directional detection experiment.  Due to the small track length, these measurements are a big experimental challenge.  Encouragingly, many directional detection experiments have published an experimentally measured angular resolution and the sense recognition threshold\,\cite{Battat:2016pap}.  More experimental work is needed to demonstrate that the angular resolution is measurable and the sense recognition is possible at lower recoil energies. 

The angular size of the ring depends on the target and the dark matter velocity distribution.  Heavier dark matter particles produce rings with bigger angular size.  We tabulate the range in the ring sizes for the two targets and the cosmological dark matter velocity distribution for halos E9 and E11 in Table\,\ref{tab:ring size} for dark matter particle masses between 100 GeV and 1 TeV.  We find that a heavier target produces a larger ring size.

\begin{table}[b]

\caption{Range of the ring sizes for various targets and dark matter velocity distributions.  The dark matter particle masses considered vary between 100 GeV and 1 TeV.}

\begin{ruledtabular}

\begin{tabular}{lccc}
  & SHM & E9 cosmological & E11 cosmological \\ 
\hline
$^{19}$F & 38$^{\degree}$ - 44$^{\degree}$ & 40$^{\degree}$ - 48$^{\degree}$ & 42$^{\degree}$ - 51$^{\degree}$\\
Xe & 44$^{\degree}$ - 61$^{\degree}$ & 46$^{\degree}$ - 72$^{\degree}$ & 49$^{\degree}$ - 83$^{\degree}$\\

\end{tabular}

\end{ruledtabular}
\label{tab:ring size}
\end{table}

\section{Results}
\label{sec:results}

\begin{figure}
[!thpb]
\centering
\includegraphics[angle=0.0,width=0.49\columnwidth]{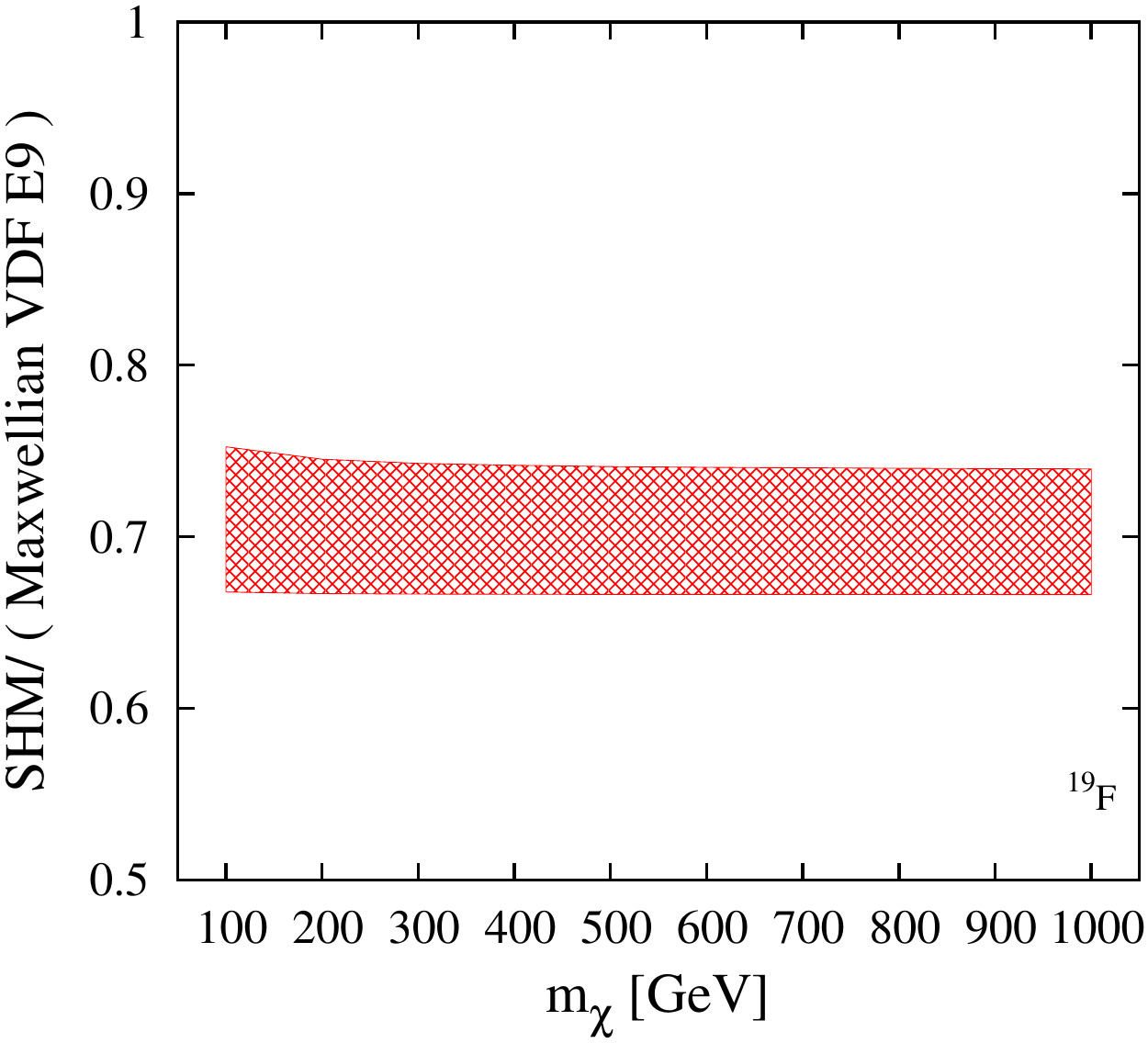}
\includegraphics[angle=0.0,width=0.49\columnwidth]{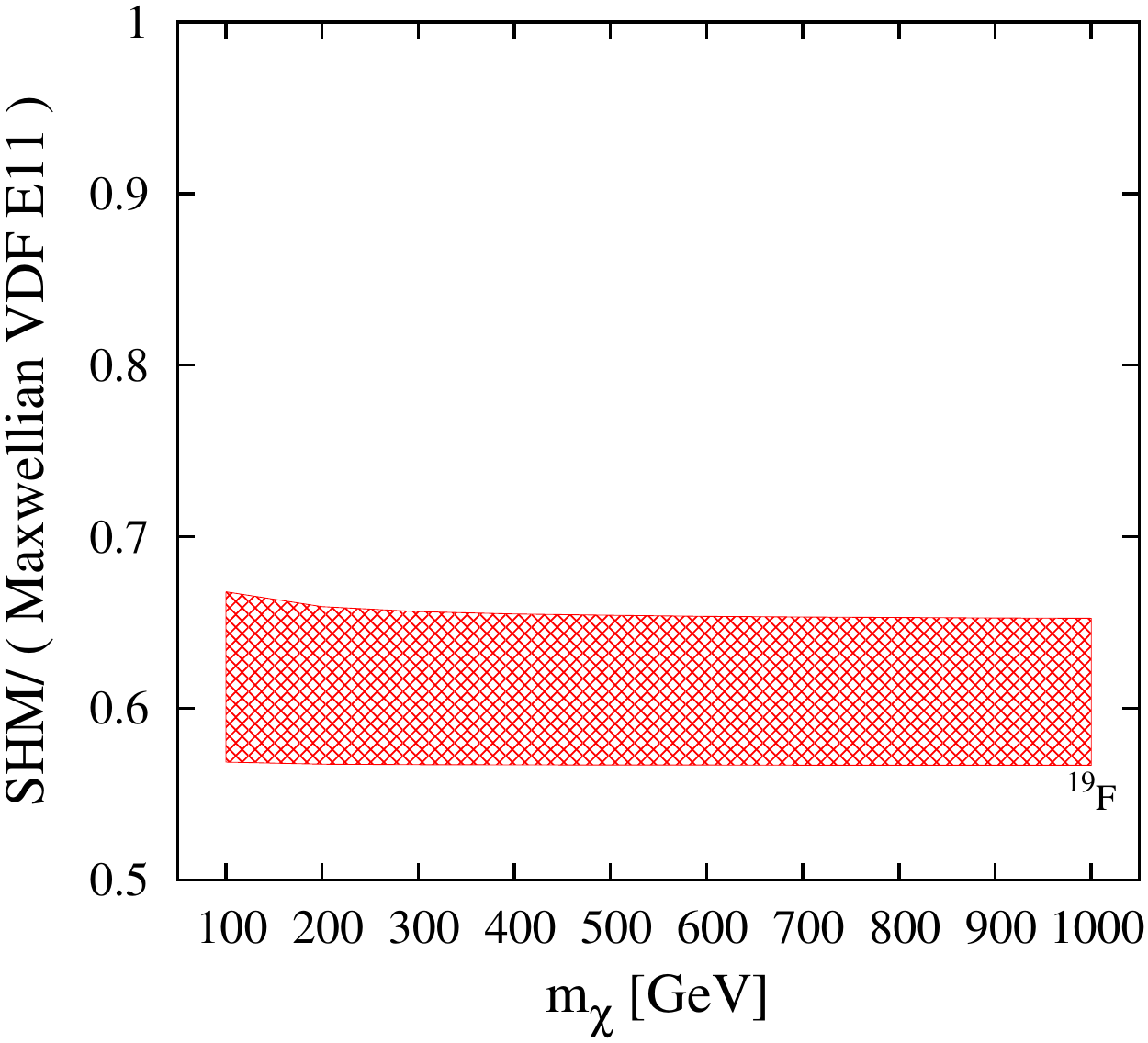}
\includegraphics[angle=0.0,width=0.49\columnwidth]{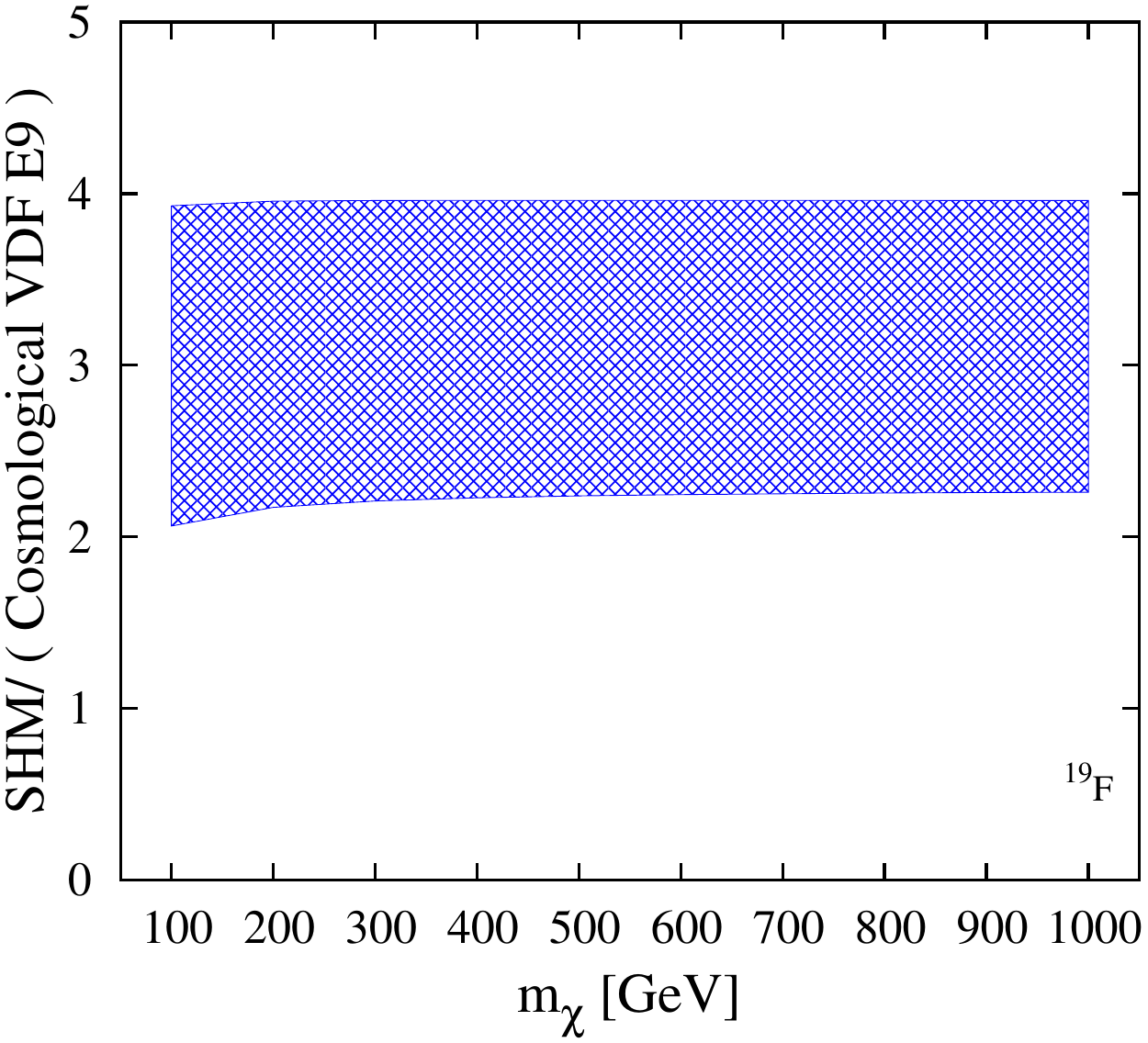}
\includegraphics[angle=0.0,width=0.49\columnwidth]{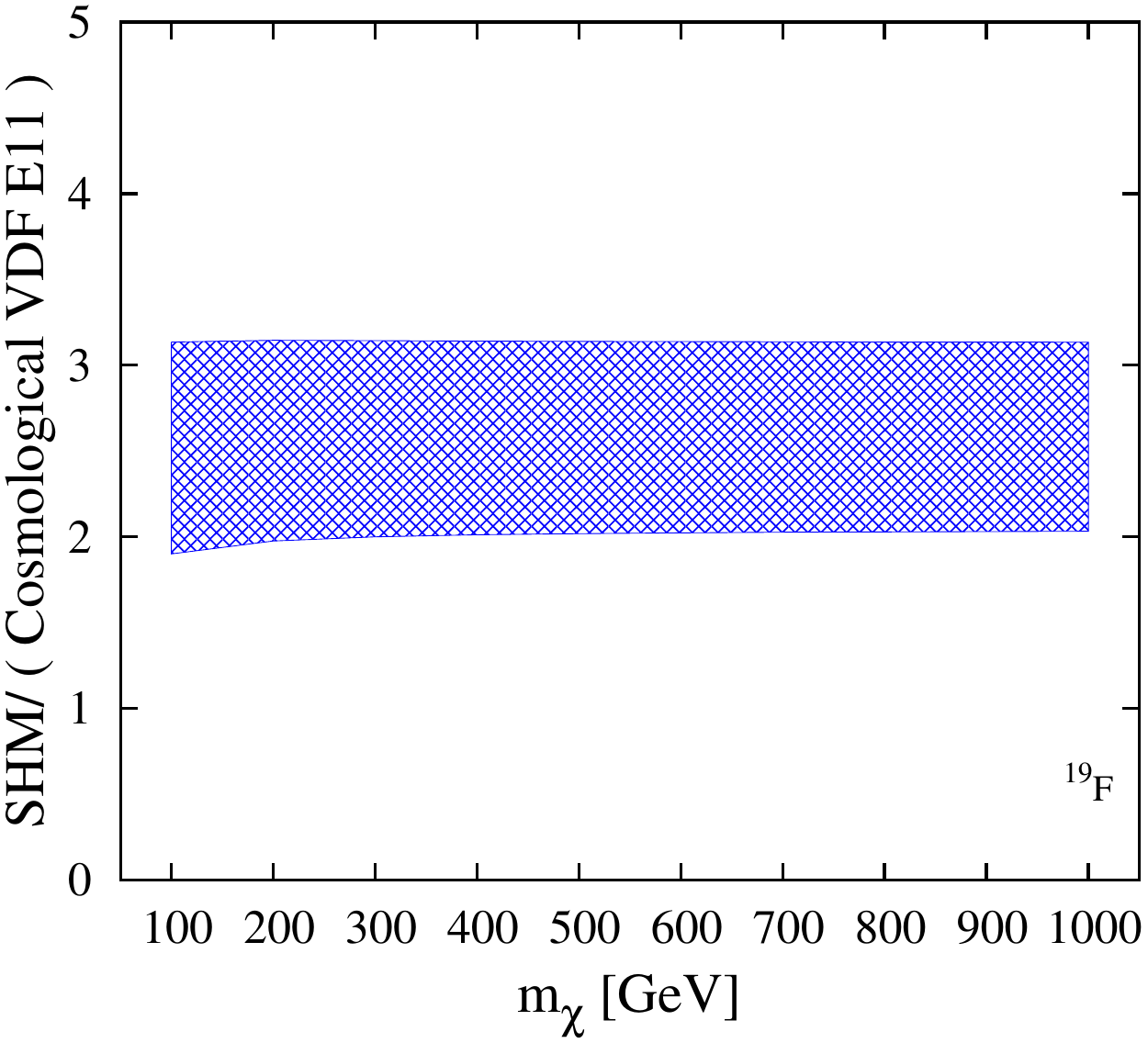}
\caption{Ratio of the number of events required (using different dark matter velocity profiles) for a 3$\sigma$ forward-backward discrimination using $^{19}$F target for various dark matter masses.  The various ratios are (clockwise from top left): $(i)$ SHM to Maxwellian VDF fit to E9, $(ii)$ SHM to Maxwellian VDF fit to E11, $(iii)$ SHM to Cosmological VDF fit to E11, and $(iv)$ SHM to Cosmological VDF fit to E9.}
\label{fig:forward backward 19F}
\end{figure}

\begin{figure}
[!thpb]
\includegraphics[angle=0.0,width=0.49\columnwidth]{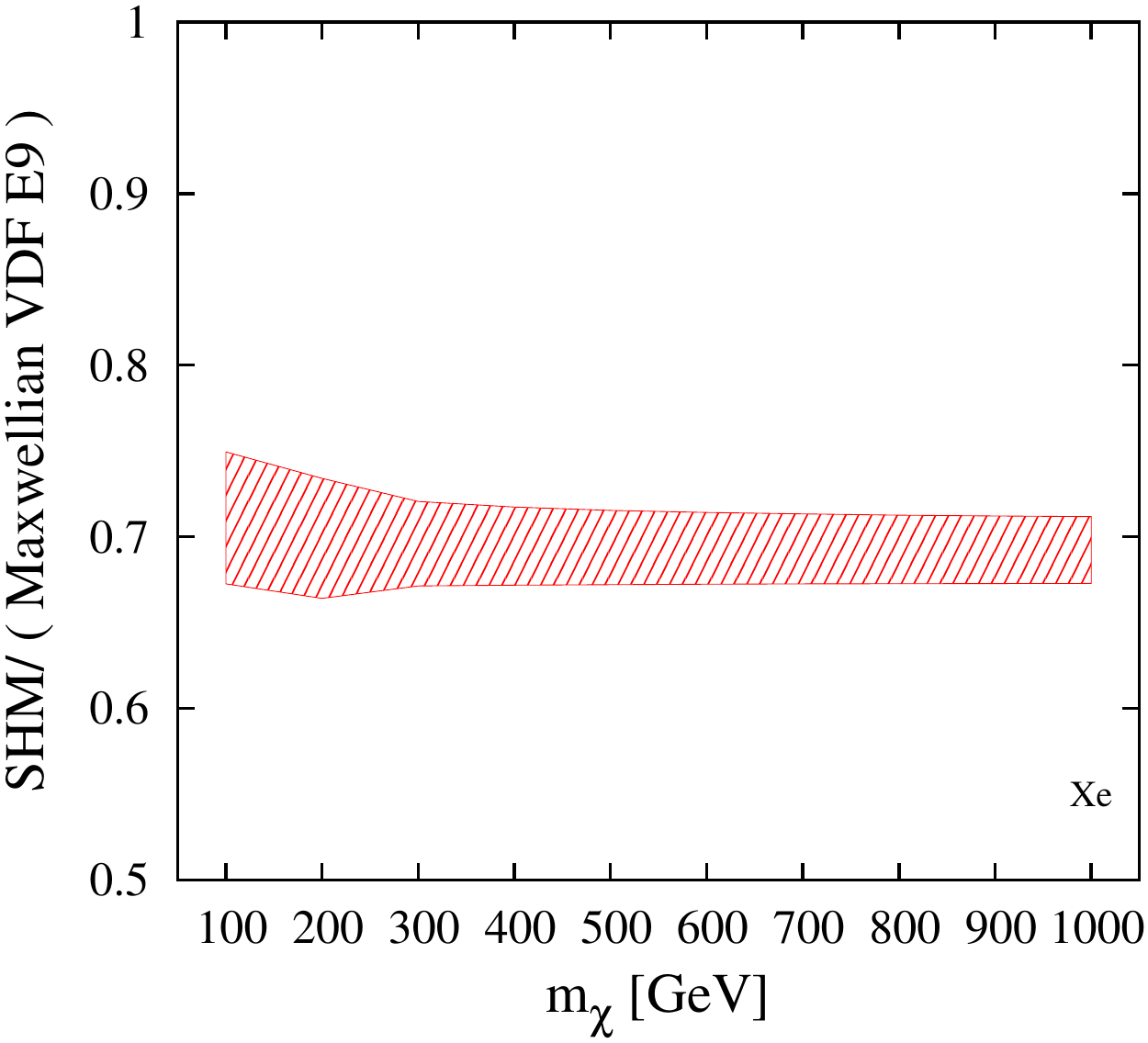}
\includegraphics[angle=0.0,width=0.49\columnwidth]{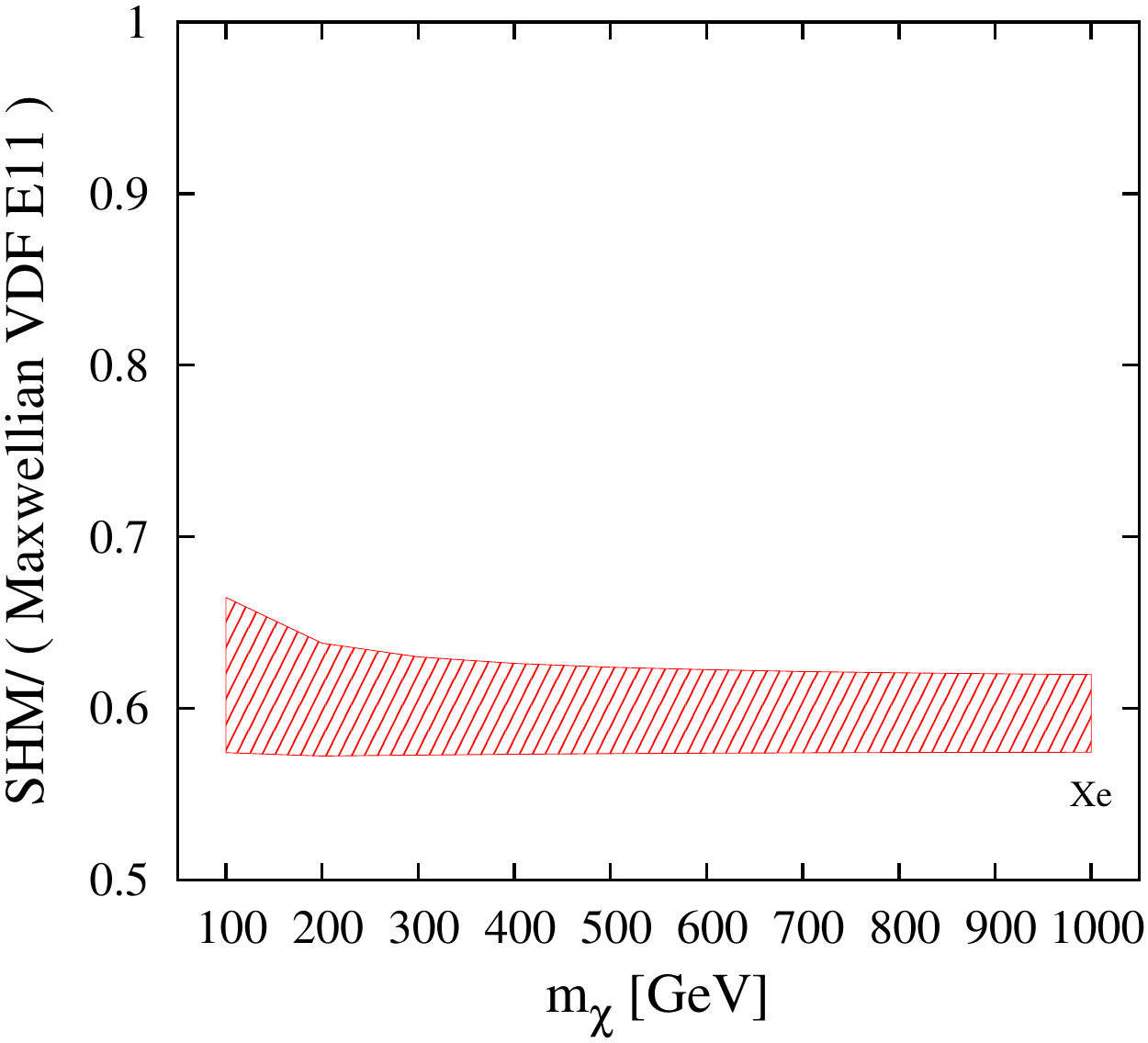}
\includegraphics[angle=0.0,width=0.49\columnwidth]{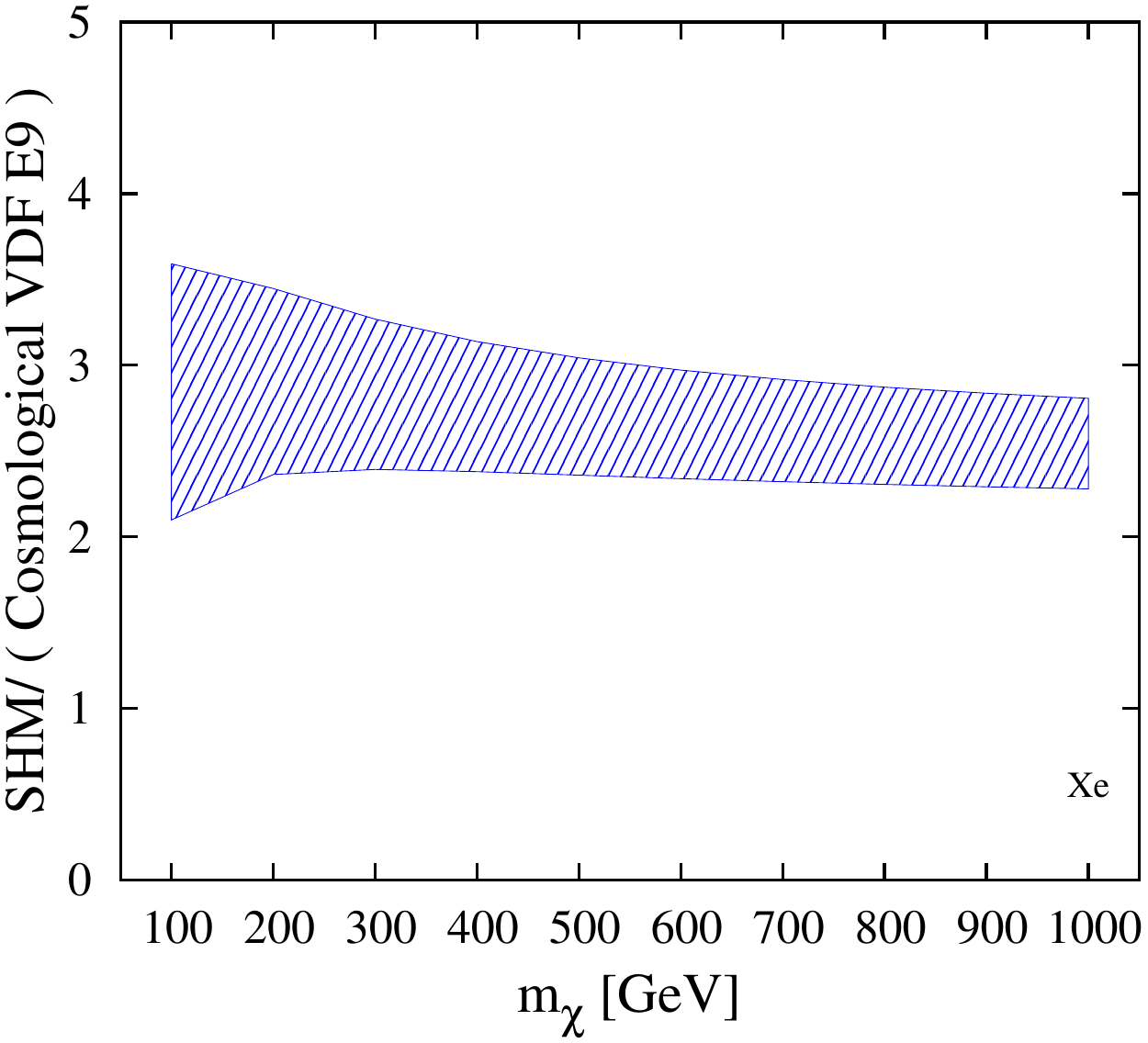}
\includegraphics[angle=0.0,width=0.49\columnwidth]{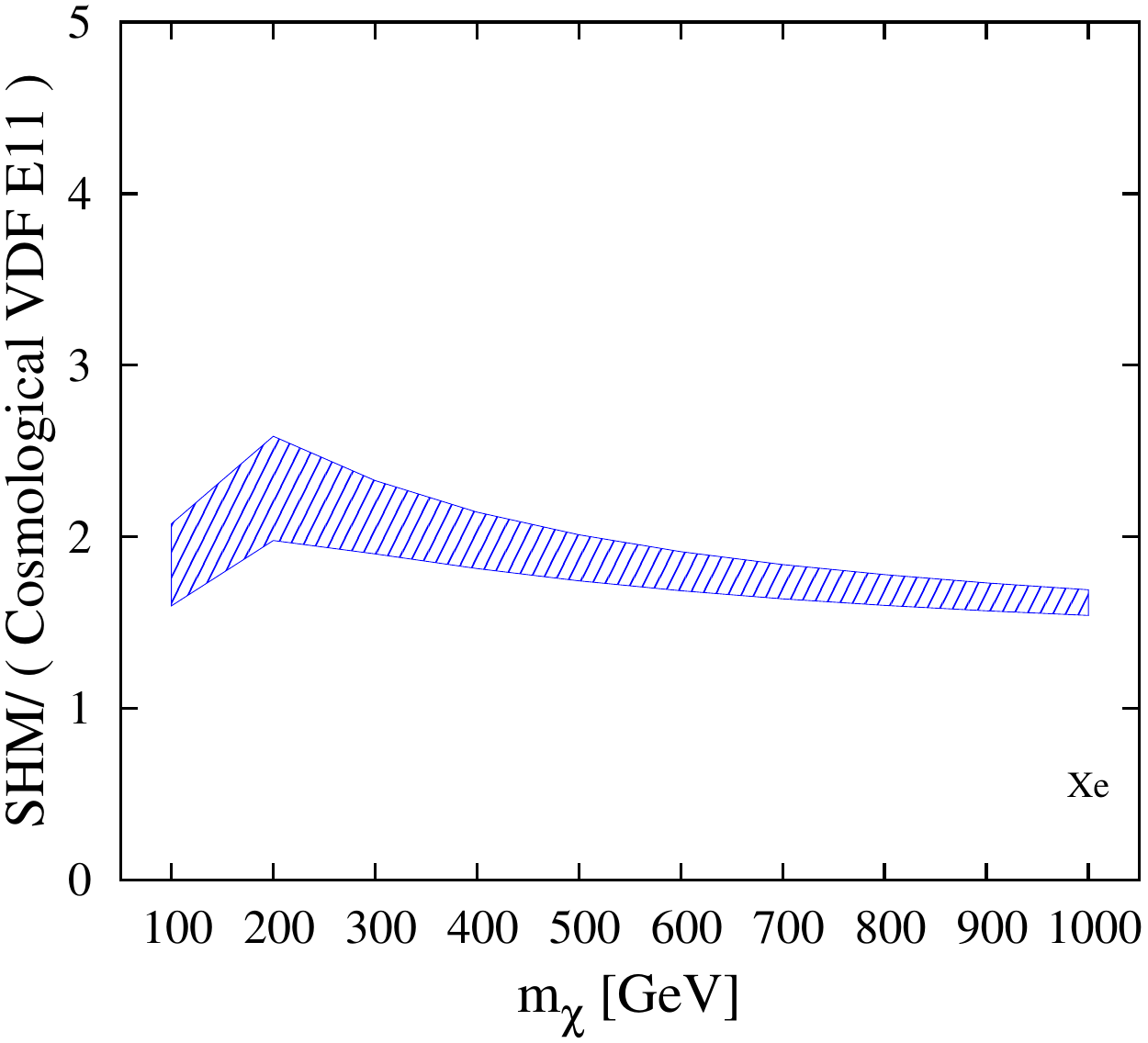}
\caption{Same as Fig.\,\ref{fig:forward backward 19F} but for Xe.}
\label{fig:forward backward Xe}
\end{figure}


In this section, we first show the nuclear recoil energy spectrum and then estimate the ratio of events to identify the forward-backward asymmetry and the ring at the 3$\sigma$ level.  

In Fig.\,\ref{fig:differential recoil rate}, we show the differential angular recoil rate when a dark matter particle of mass 100 GeV collides with a $^{19}$F and a Xe nucleus.  The differential angular recoil rate when the dark matter velocity distribution follows the Mao et al. distribution are shown by solid blue and magenta lines.  The differential angular recoil rate when the dark matter velocity distribution follows the standard Maxwellian distribution are shown by dotted red and brown lines.  The differential angular recoil rate for the SHM is shown by the black dotted line.  We integrate over the energy range [5, 10] keV for the $^{19}$F target, and over the energy range [5, 20] keV for the Xe target.  Our choice of the energy range maximizes the contrast of the ring.  A wider energy range will increase the number of recoil events in the forward direction and thus the distinct bump-like feature of the ring is washed out.  The local dark matter density is taken to be 0.3 GeV/ cm$^3$, and the spin-dependent dark matter - nucleon cross section is taken to be 10$^{-40}$ cm$^2$ for both these figures.  The angular recoil spectrum is directly proportional to this cross section and a smaller value will decrease it proportionately.  The nuclear form factor is taken from Ref.\,\cite{Bednyakov:2006ux}.  

The forward backward asymmetry is clearly visible for all the velocity profiles.  The Mao et al. profile shows the most dramatic forward-backward asymmetry.  The forward-backward asymmetry for the SHM and the standard Maxwellian distribution is weaker.  This can be easily understood from Eqn.\,\ref{eq:forward-backward ratio}.  The Mao et al. profile has a smaller ``effective $v_0$", and hence a larger forward-backward asymmetry.  The SHM and the standard Maxwellian have much larger $v_0$, and hence the contrast in their forward-backward asymmetry is much smaller. 

The angle $\theta_{v_Eq}$ at which the differential angular recoil rate is maximized is called the ``ring" angle.  As analytically explained in Eqn.\,\ref{eq:ring contrast}, the Mao et al. profile fit produces the largest ring contrast, whereas the SHM and the standard Maxwellian profile fit produces a much weaker ring contrast.

We perform a simple statistical test\,\cite{Bozorgnia:2011vc} to determine the ratio of the number of events required for 3$\sigma$ discrimination for the forward-backward asymmetry for different dark matter velocity profiles.  We also calculate the ratio of events for a 3$\sigma$ discovery of the ring for various different dark matter profiles.  Using the ratio of events makes our results independent of the local dark matter density, dark matter - nucleon cross section, and many other uncertainties.

We briefly describe the procedure that we follow, and then describe the results.  We calculate the number of events in the forward and backward direction by integrating over $\theta_{v_Eq} \in [\pi/2, \pi]$ and $\theta_{v_Eq} \in [0, \pi/2]$ respectively.  We construct the forward-backward asymmetry as (N$_{\rm F}$ - N$_{\rm B}$)/$\sqrt{{\rm N}_{\rm F} + {\rm N}_{\rm B}}$.  We increase the exposure so that (N$_{\rm F}$ - N$_{\rm B}$)/$\sqrt{{\rm N}_{\rm F} + {\rm N}_{\rm B}}$ = 3.  For this exposure, we calculate the total number of events for the given velocity distribution.

As expected from Fig.\,\ref{fig:differential recoil rate}, the 3$\sigma$ discrimination in the forward-backward asymmetry is achieved with a smaller number of events for the Mao et al. velocity profile, whereas the standard Maxwellian fit to the halos E9 and E11 require the largest number of events for this discrimination.  Fig.\,\ref{fig:forward backward 19F} shows the ratio of the number of events for various different input dark matter velocity profile required for such a discrimination with a $^{19}$F target.  In the top panel, we plot the ratio of the number of events required in the SHM to that of the standard Maxwellian fit to halos E9 and E11 for various dark matter masses.  The width of the bands are calculated taking the Poisson uncertainty in both the numerator and the denominator.  It can be seen that the number of events required in the SHM is $\sim$60\% - 70\% of that required in the standard Maxwellian velocity distribution function.  

In the lower panel of Fig.\,\ref{fig:forward backward 19F}, we show the ratio of the events required for the SHM velocity profile to the Mao et al. fit to the halos E9 and E11.  Since the ring contrast is much larger for the Mao et al. fit to these halos, the number of events required for this discrimination is $\sim$2 -- 3 times smaller than that required in SHM.  

In Fig.\,\ref{fig:forward backward Xe}, we show the same for Xe target.  Even in this case, the Mao et al. fit to halos E9 and E11 require the least number of events to achieve the 3$\sigma$ discrimination in the forward-backward asymmetry.  The bands representing the ratio has a smaller width compared to that of the $^{19}$F target.  This is because the total number of events required for the 3$\sigma$ discrimination for Xe target is larger than that of $^{19}$F target.

We define the ring following Ref.\,\cite{Bozorgnia:2011vc}.  The ring is defined to be between angles $\theta_{v_Eq1} < \theta_{v_Eq} < \theta_{v_Eq2}$, where
\begin{eqnarray}
\dfrac{dR}{d\Omega_{v_Eq1}} = \dfrac{dR}{d\Omega_{v_Eq2}} &=& \dfrac{1}{2} \, \Bigg(\dfrac{dR}{d\Omega_{v_Eq}}(\theta_{v_Eq} = \pi) \nonumber\\
&+&\, \dfrac{dR}{d\Omega_{v_Eq}}\Bigg|_{\rm max} \Bigg) \, .
\label{eq:ring}
\end{eqnarray}
For these two angles, $\theta_{v_Eq1}$ and $\theta_{v_Eq2}$, we calculate the number of events inside these angles and between $\theta_{v_Eq2}$ and $\pi$.  We define $N_{12} = \int_{\theta_{vEq1}} ^{\theta_{v_Eq2}} d\Omega  \, \dfrac{dR}{d\Omega_{v_Eq}}$, and $N_{2\pi} = \int_{\theta_{vEq2}} ^\pi d\Omega  \, \dfrac{dR}{d\Omega_{v_Eq}}$.  We calculate the exposure required such that $(N_{12} - N_{2\pi})/\sqrt{N_{12} + N_{2\pi}}$ = 3.  For such an exposure, we calculate the total number of events in the energy range [5, 10] keV for $^{19}$F and [5, 20] keV for Xe.  We find that using a factor of $1/\sqrt{2}$ or $1/3$ in eqn.\,\ref{eq:ring} produces either a ring which is too small to be detected due to limitations imposed by angular resolution or a ring which is so thick that larger number of events are required for discovery. 

\begin{figure}
\includegraphics[angle=0.0,width=0.48\columnwidth]{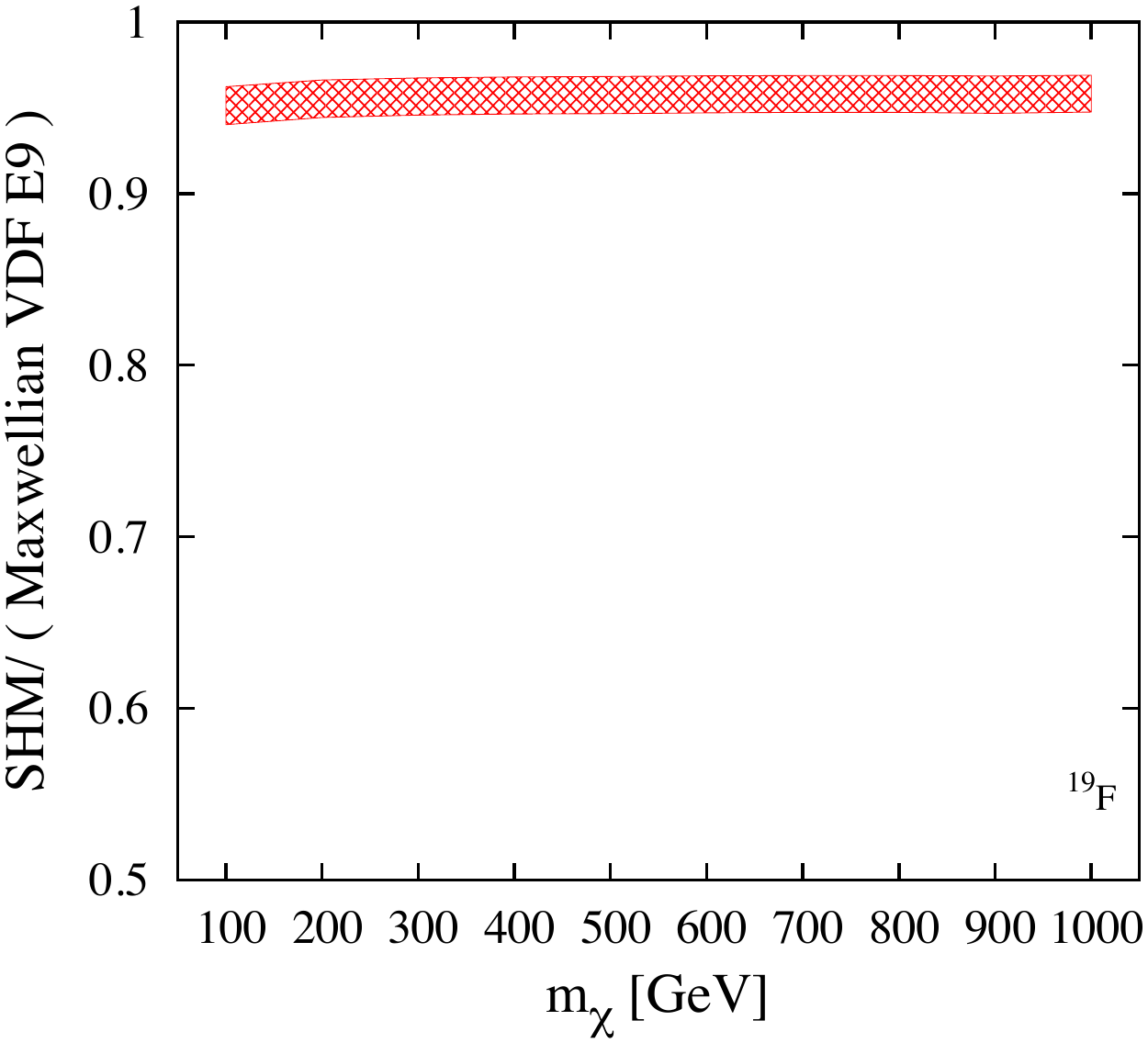}
\includegraphics[angle=0.0,width=0.48\columnwidth]{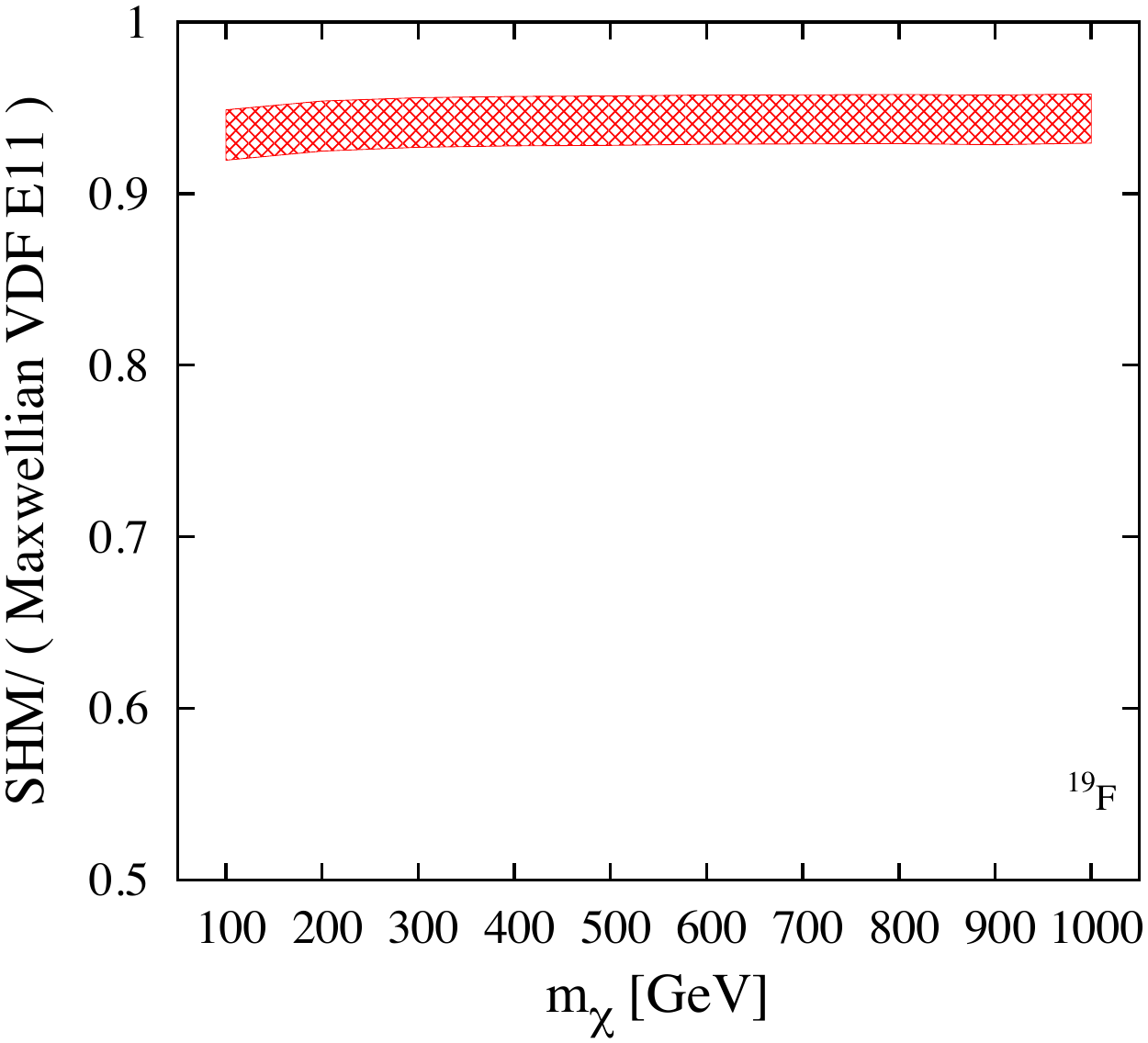}
\includegraphics[angle=0.0,width=0.48\columnwidth]{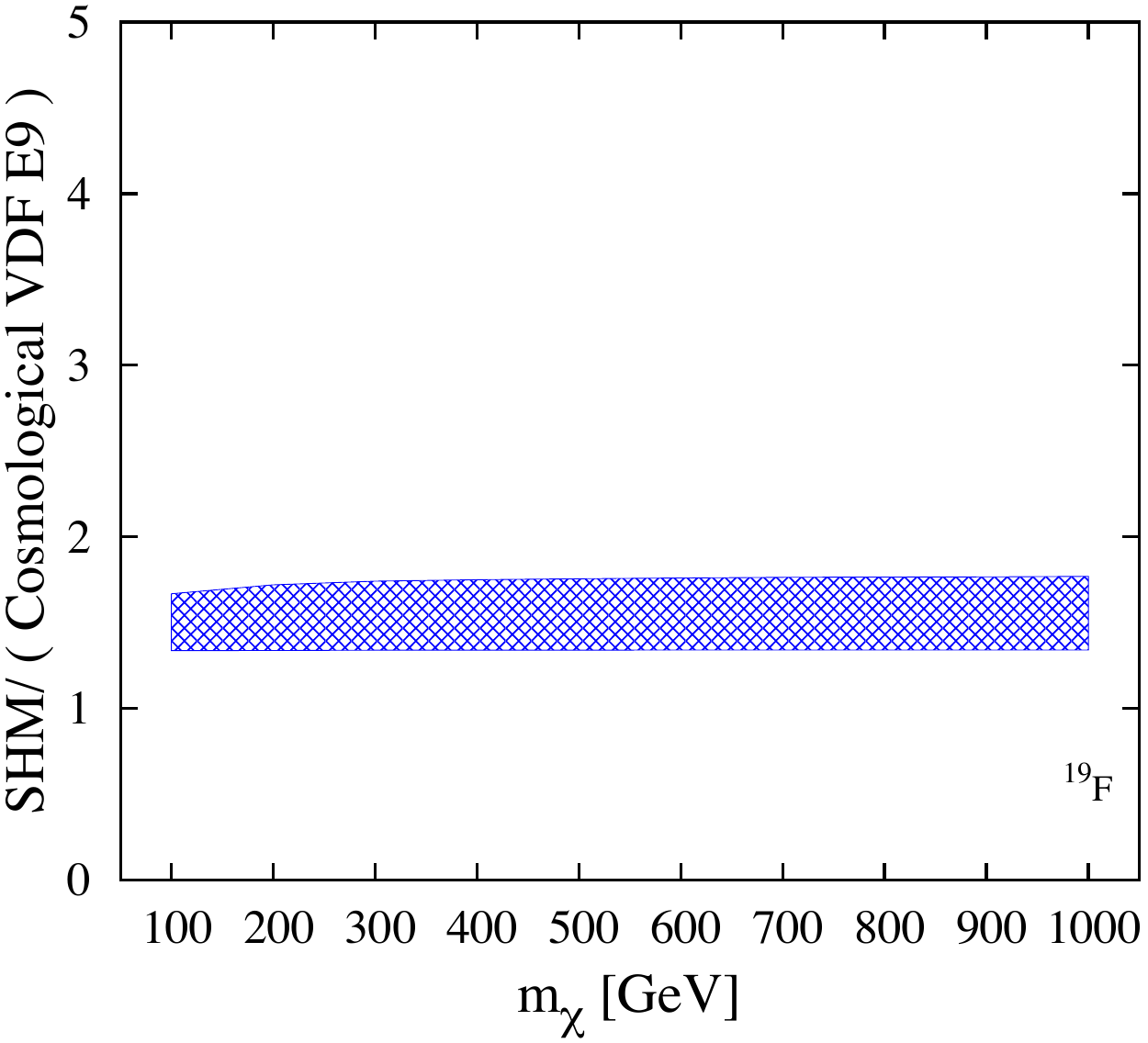}
\includegraphics[angle=0.0,width=0.48\columnwidth]{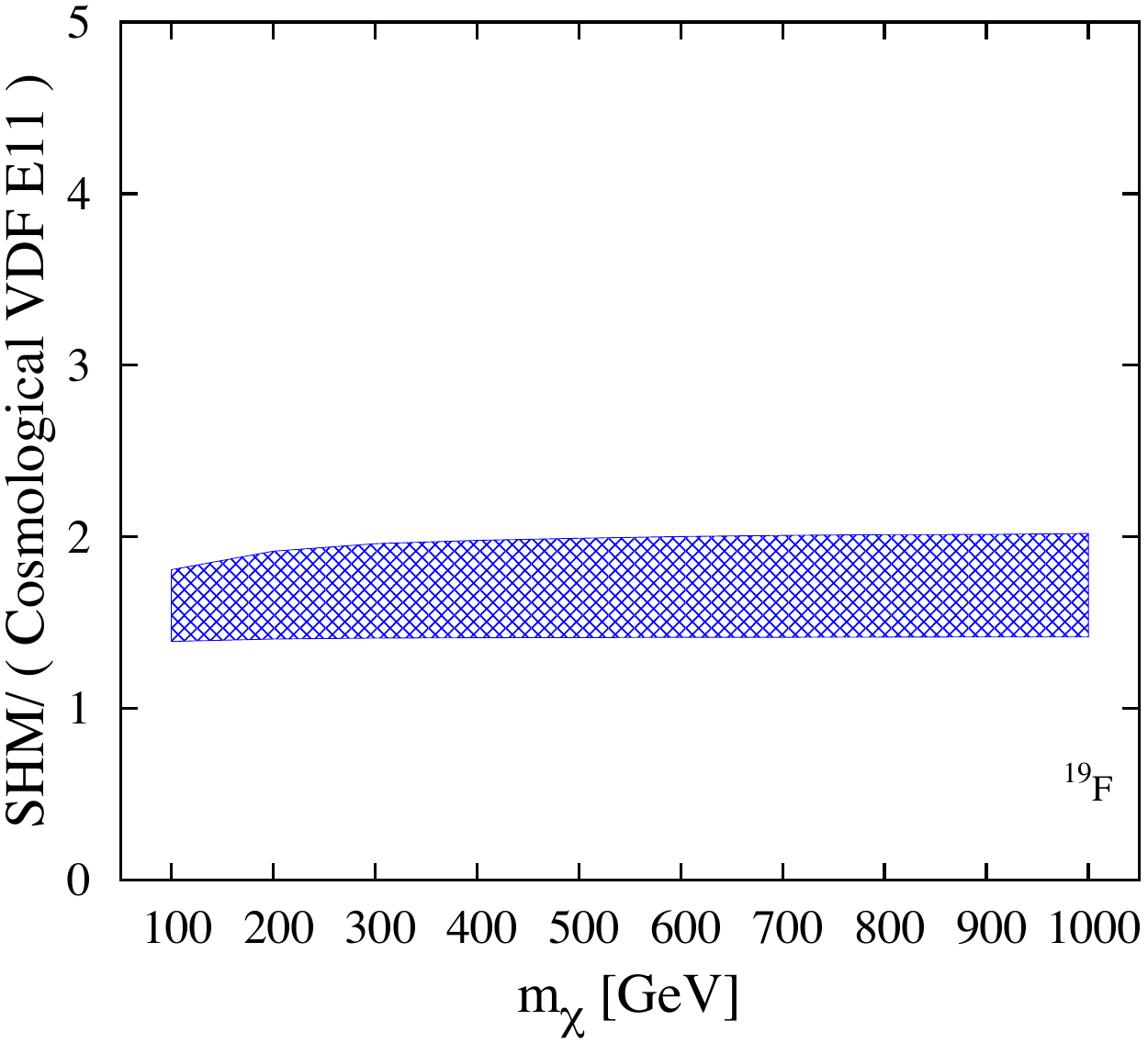}
\caption{Ratio of the number of events required (using different dark matter velocity profiles) for a 3$\sigma$ evidence of a ring using $^{19}$F target for various dark matter masses.  The various ratios are (clockwise from top left): $(i)$ SHM to Maxwellian VDF fit to E9, $(ii)$ SHM to Maxwellian VDF fit to E11, $(iii)$ SHM to Cosmological VDF fit to E11, and $(iv)$ SHM to Cosmological VDF fit to E9.}
\label{fig:ring 19F}
\end{figure}

\begin{figure}
\includegraphics[angle=0.0,width=0.49\columnwidth]
{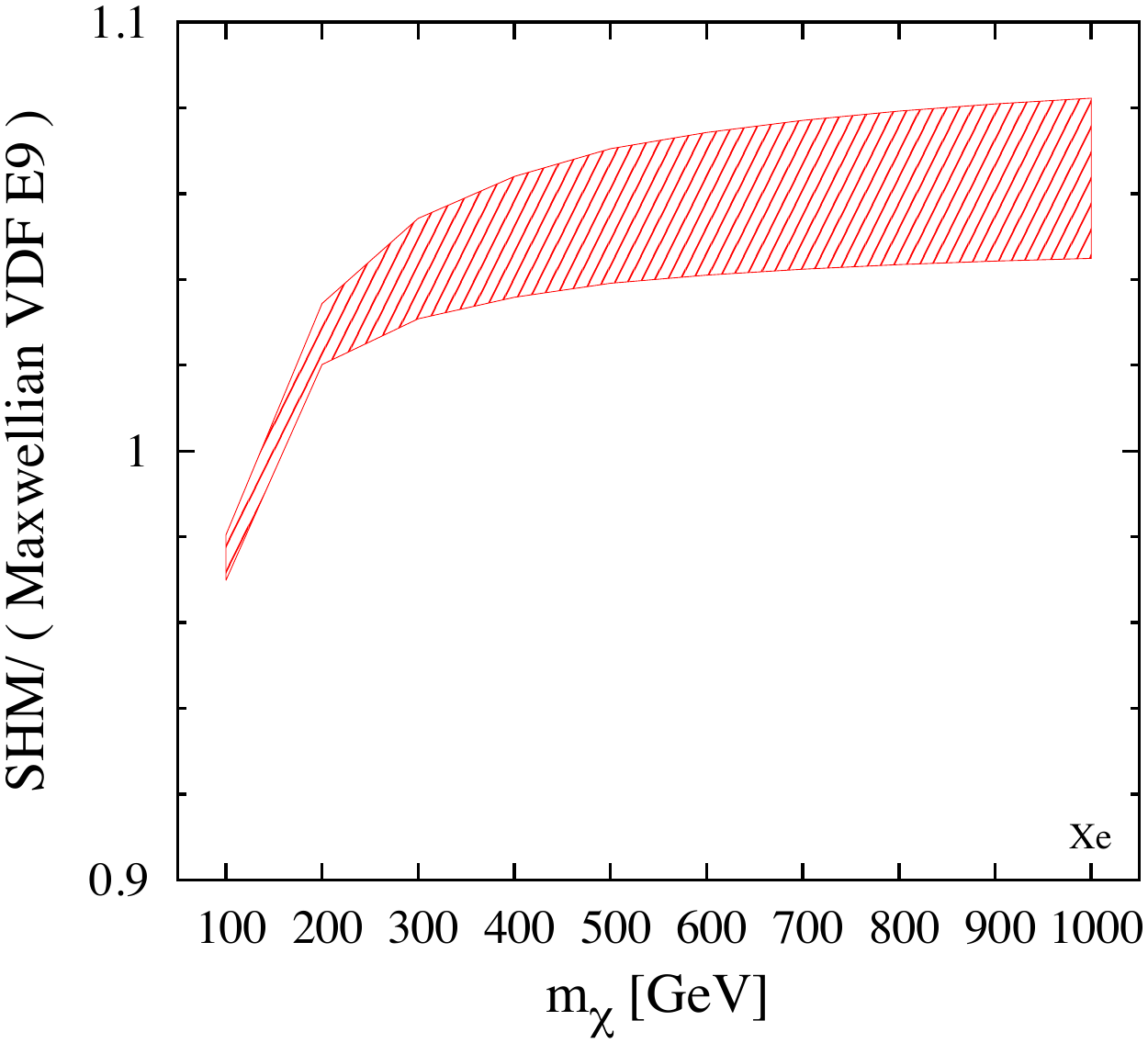}
\includegraphics[angle=0.0,width=0.49\columnwidth]{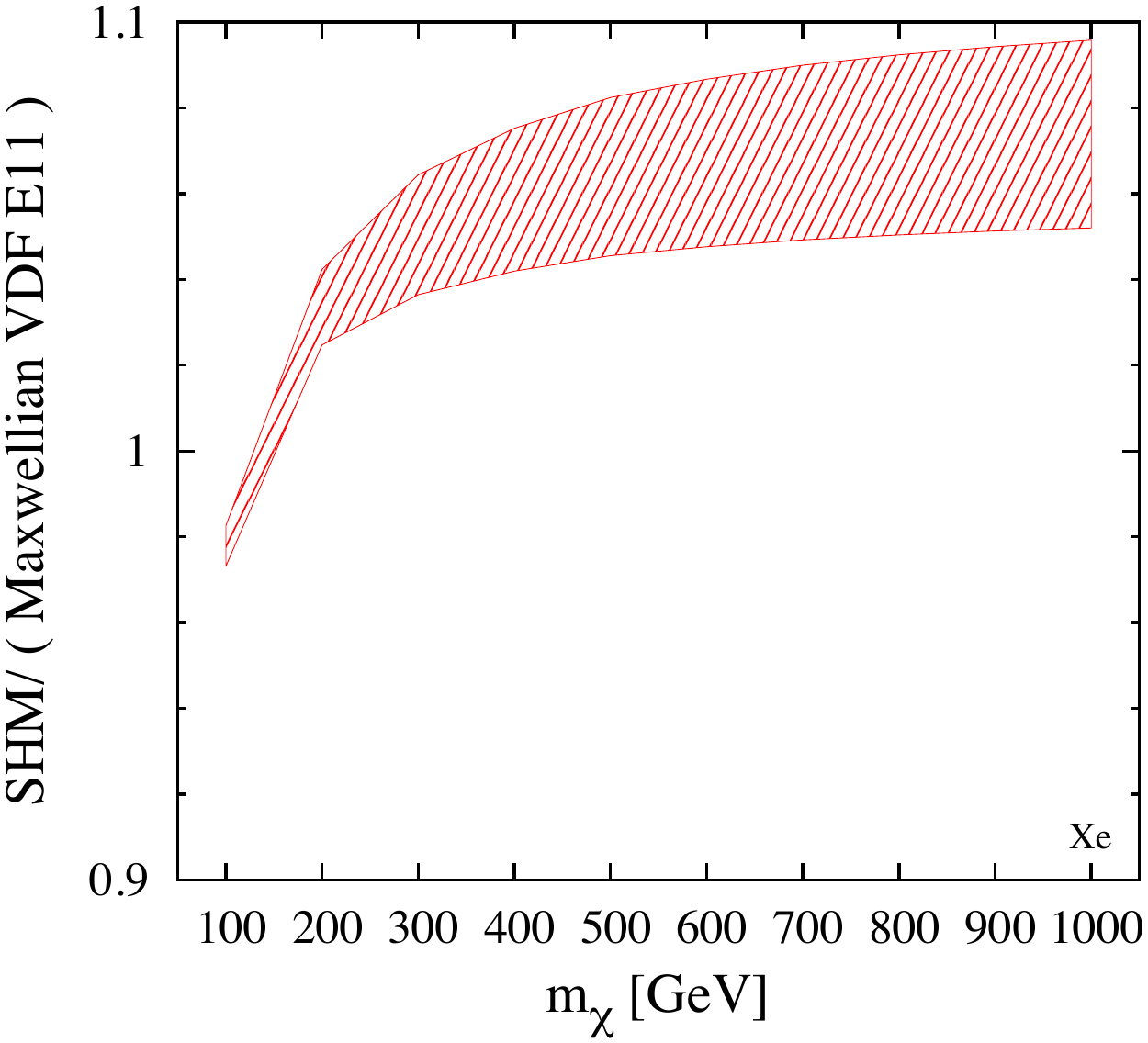}
\includegraphics[angle=0.0,width=0.49\columnwidth]{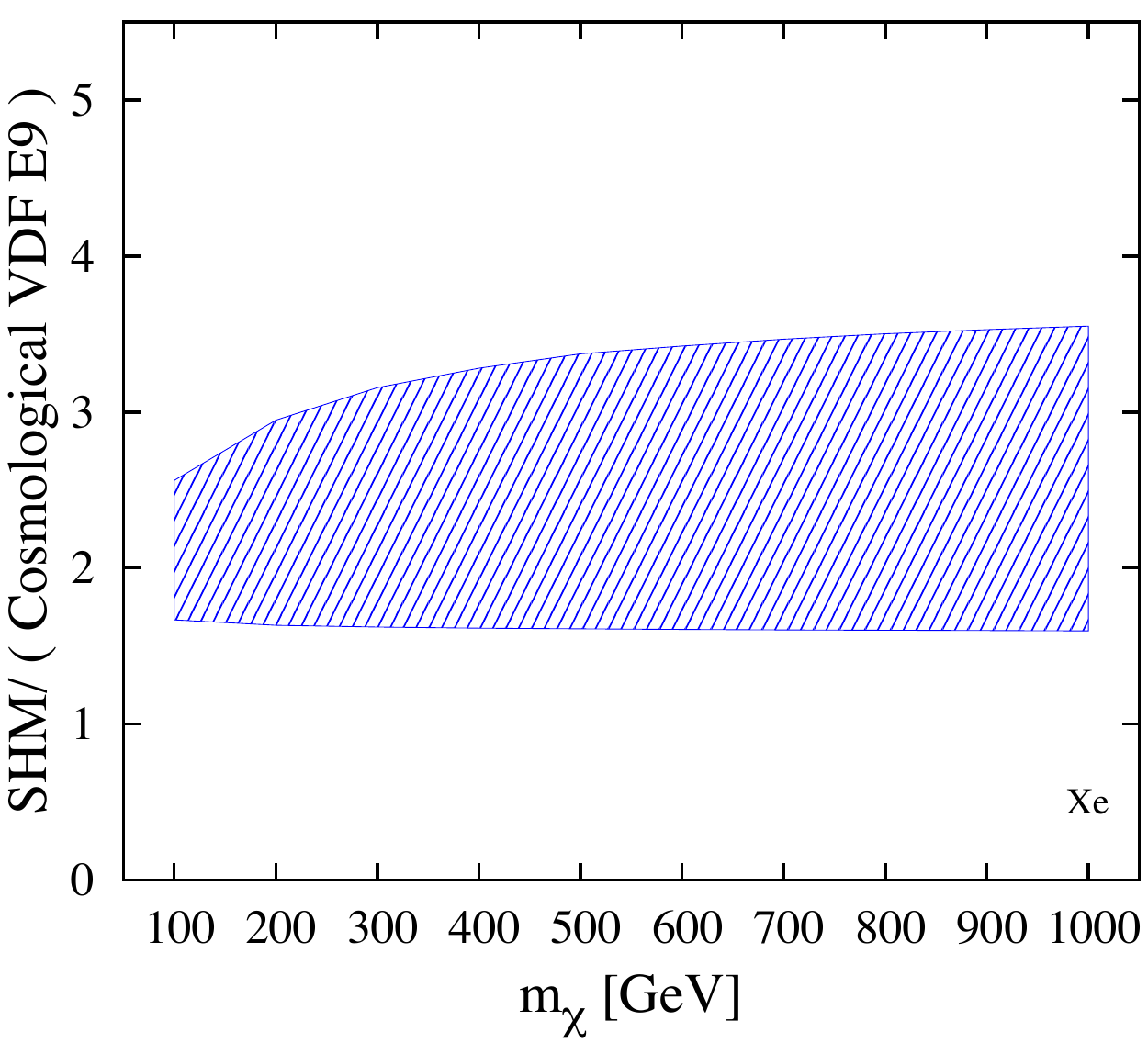}\includegraphics[angle=0.0,width=0.49\columnwidth]{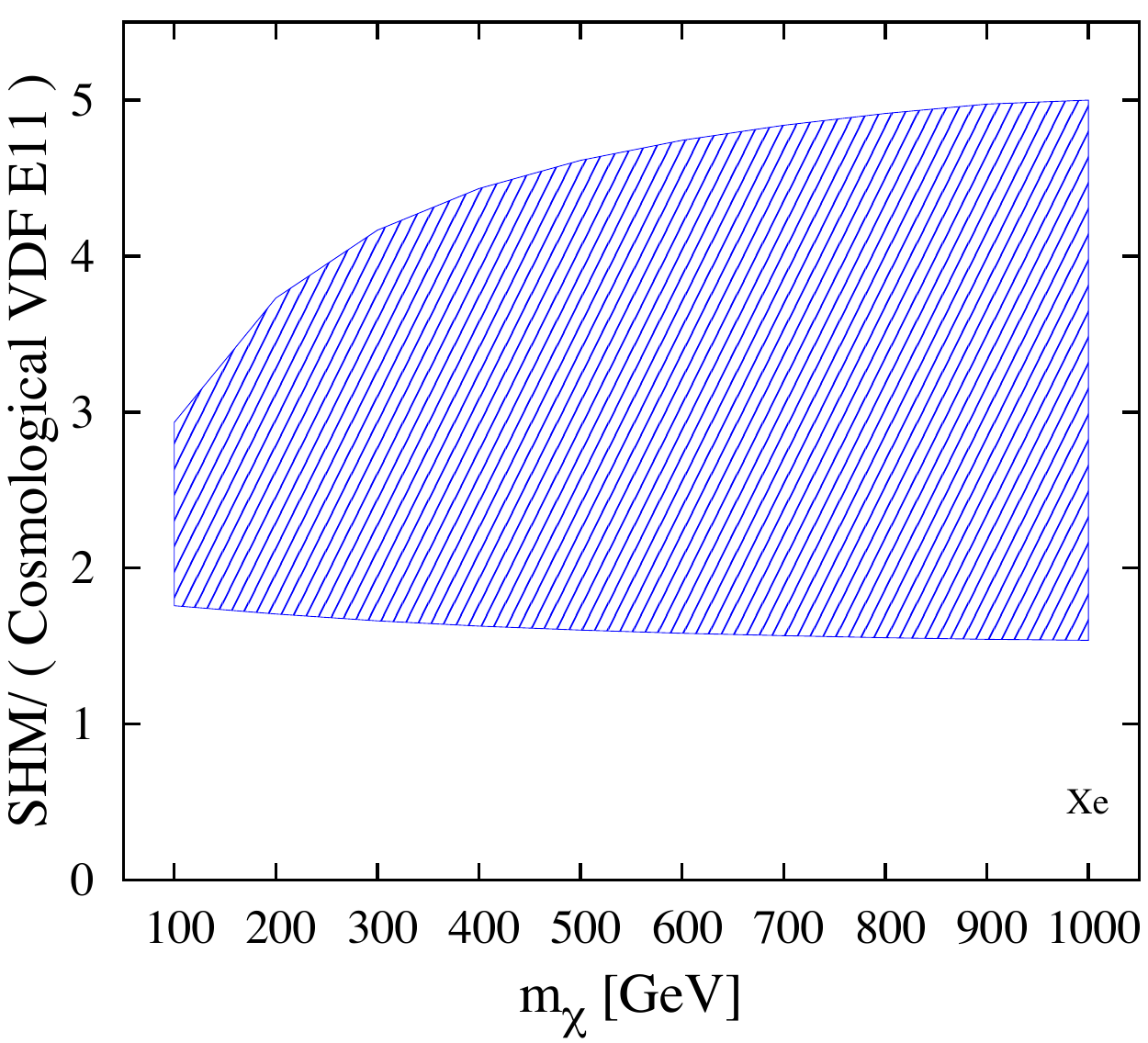}
\caption{Same as Fig.\,\ref{fig:ring 19F} but for Xe.}
\label{fig:ring Xe}
\end{figure}


The ratio of the number of events for the discrimination of the ring for various dark matter velocity profiles is shown in Figs.\,\ref{fig:ring 19F} and \ref{fig:ring Xe} for $^{19}$F and Xe target respectively.  The number of events required in the SHM for the ring discrimination (with $^{19}$F target) is very similar to the number of events required in the standard Maxwellian fit to the halos E9 and E11 (Fig.\,\ref{fig:ring 19F} top panel).  The number of events required in a $^{19}$F target for the ring discrimination in the case of the Mao et al. profile is about a factor of 2 smaller than in the SHM (Fig.\,\ref{fig:ring 19F} bottom panel).   

The corresponding figures for the Xe target is shown in Fig.\,\ref{fig:ring Xe}.  The number of events required for ring discrimination for SHM and standard Maxwellian fit to halos E9 and E11 are almost the same.  The sharp downturn of the ratio reflects the fact that the reduced mass of the dark matter - Xe nucleus system changes a lot more slowly once the dark matter mass is greater than the mass of the relevant Xe nucleus.  The number of events required for ring discrimination for the Mao et al. velocity distribution fit to halos E9 and E11 is $\sim$2 - 3 times smaller than that required for the SHM velocity distribution.

\begin{figure}
\includegraphics[angle=0.0,width=0.48\columnwidth]{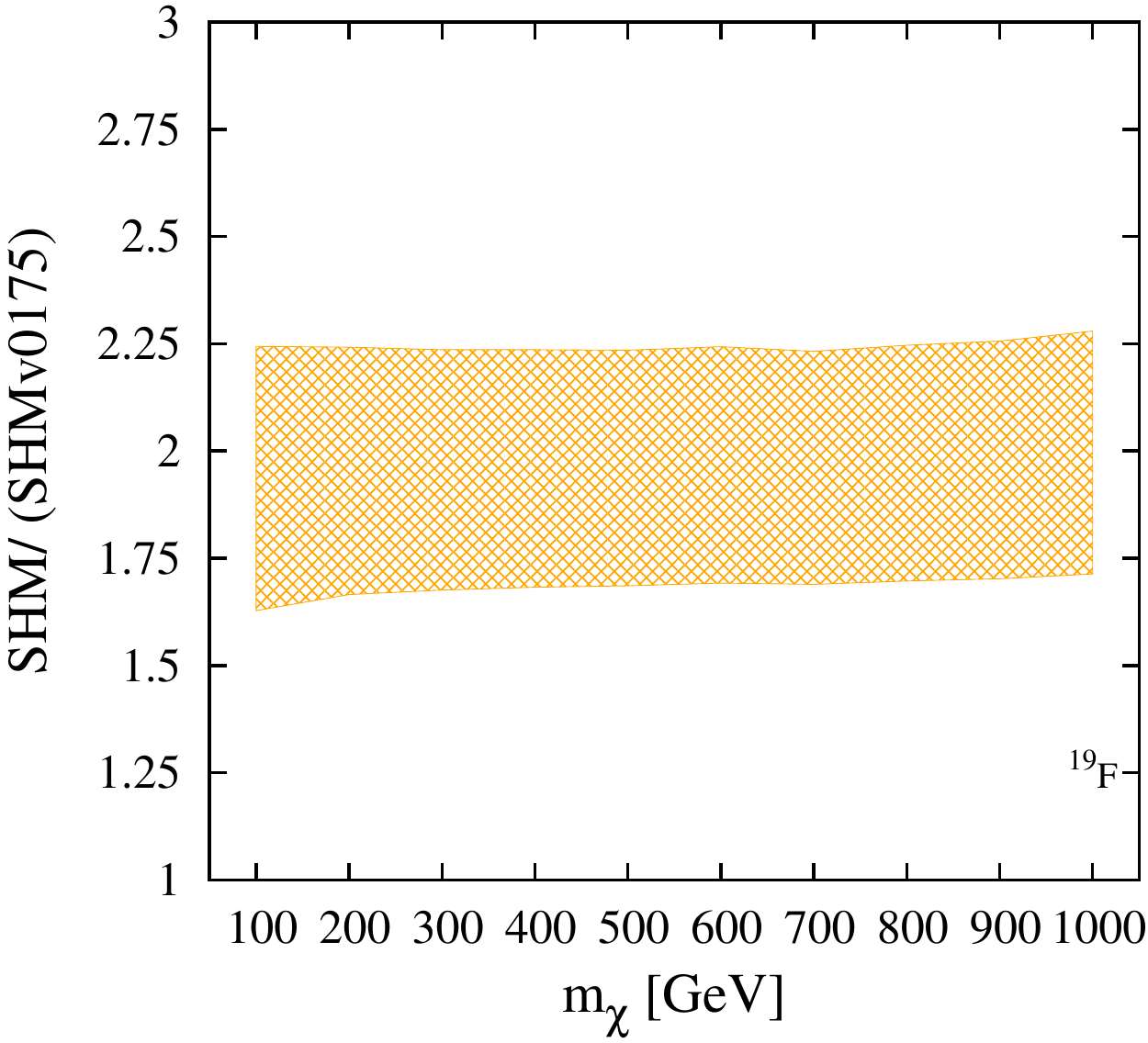}
\includegraphics[angle=0.0,width=0.48\columnwidth]{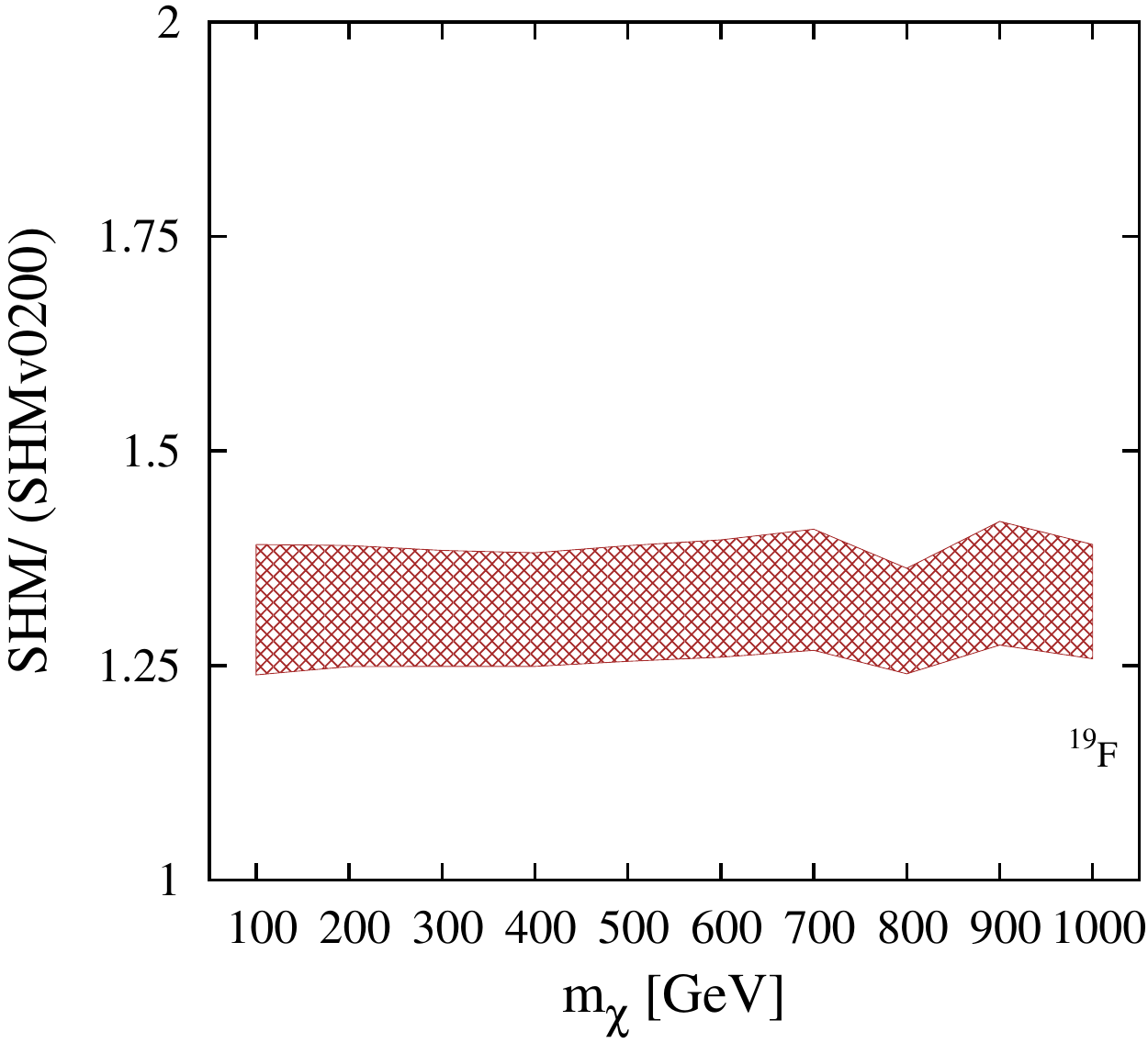}
\includegraphics[angle=0.0,width=0.48\columnwidth]{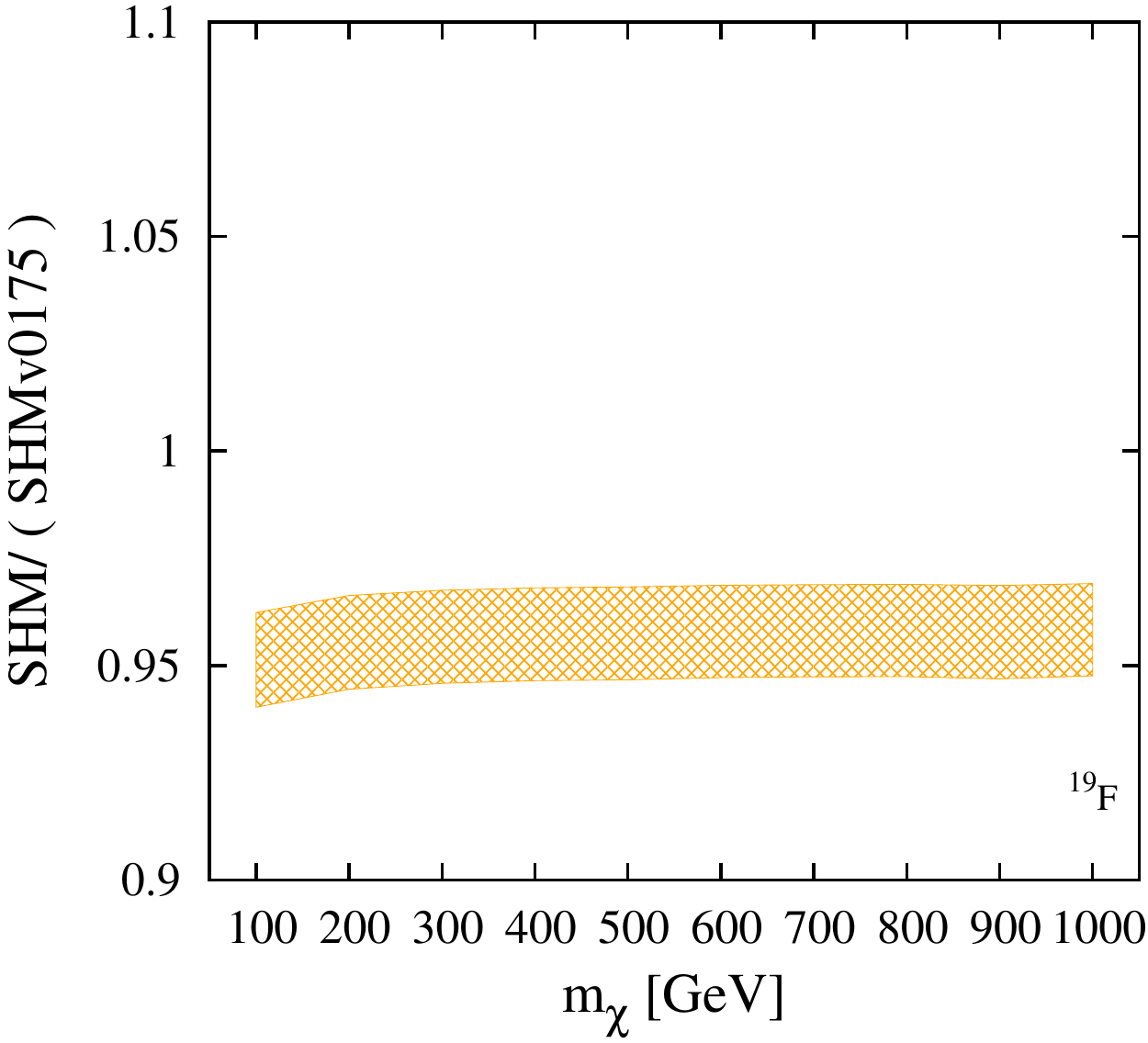}
\includegraphics[angle=0.0,width=0.48\columnwidth]{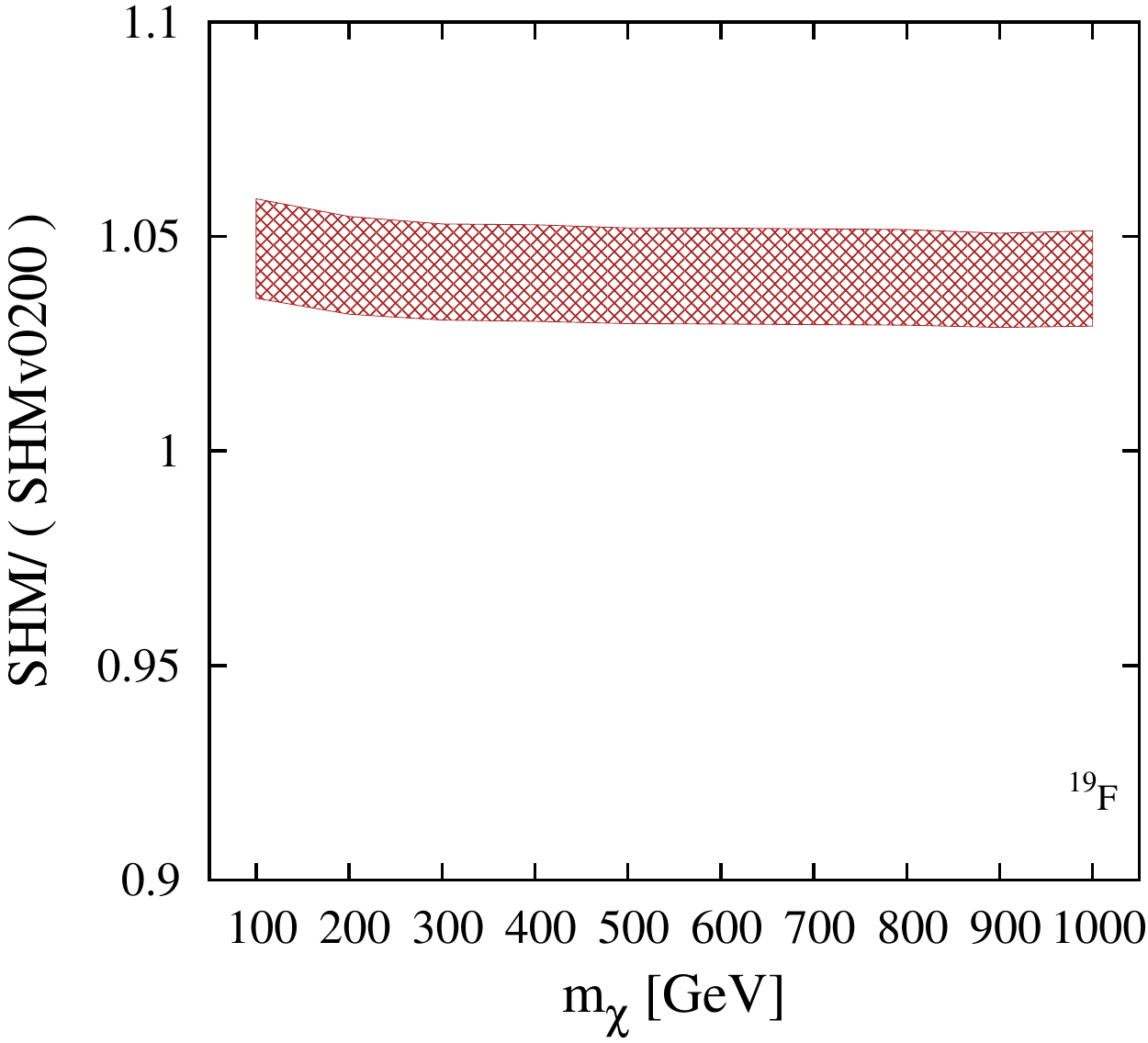}
\caption{Ratio of the number of events required (using different dark matter velocity profiles) for a 3$\sigma$ evidence of a forward - backward asymmetry (top panel) and a ring (bottom panel) using $^{19}$F target for various dark matter masses.  The ratios for the plots on the left are for SHM with $v_0 = 220$ km s$^{-1}$ to SHM with $v_0 = 175$ km s$^{-1}$.  The ratios for the plots on the right are for SHM with $v_0 = 220$ km s$^{-1}$ to SHM with $v_0 = 175$ km s$^{-1}$.}
\label{fig:v175 v200 19F}
\end{figure}

\begin{figure}
\includegraphics[angle=0.0,width=0.48\columnwidth]{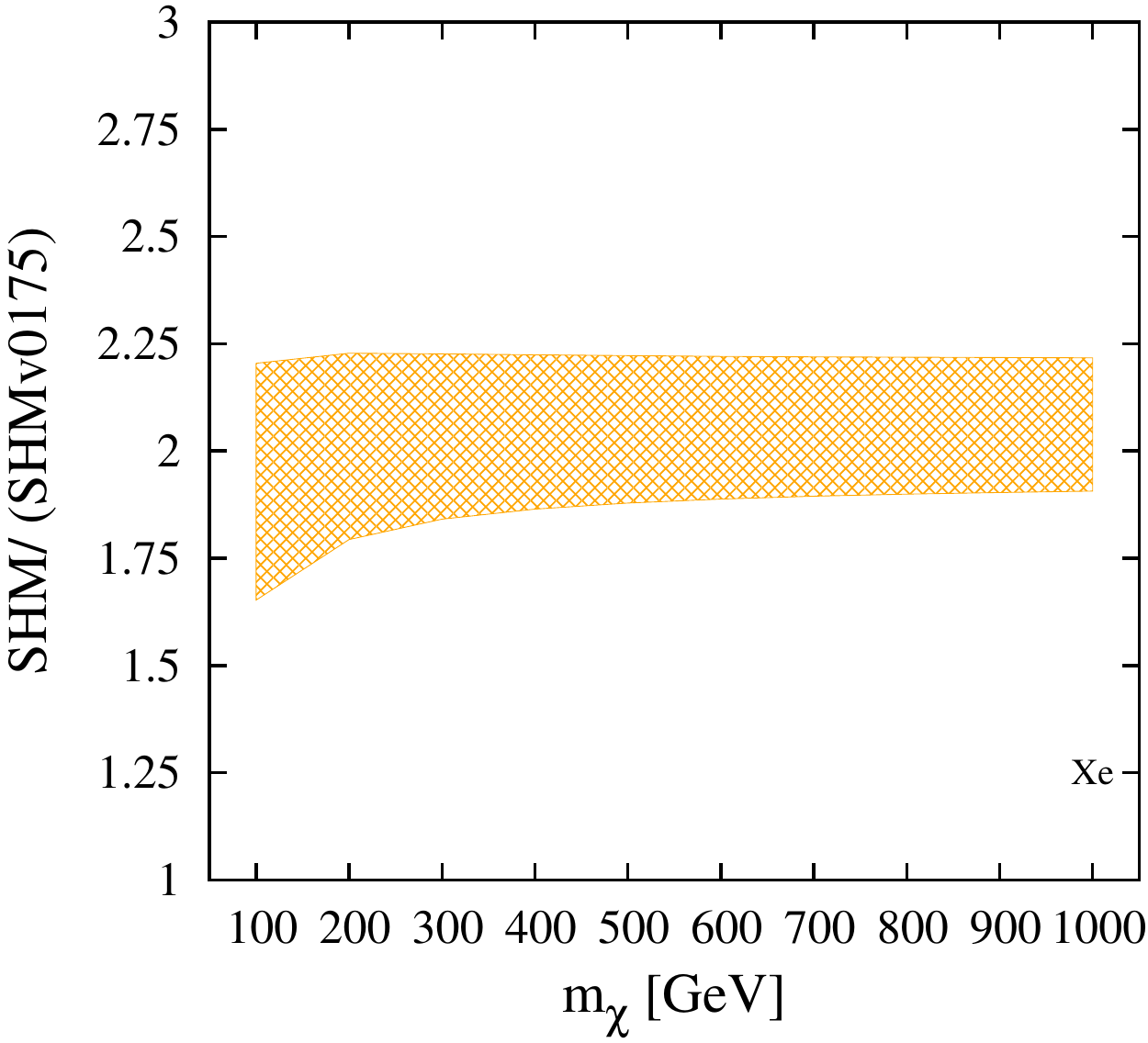}
\includegraphics[angle=0.0,width=0.48\columnwidth]{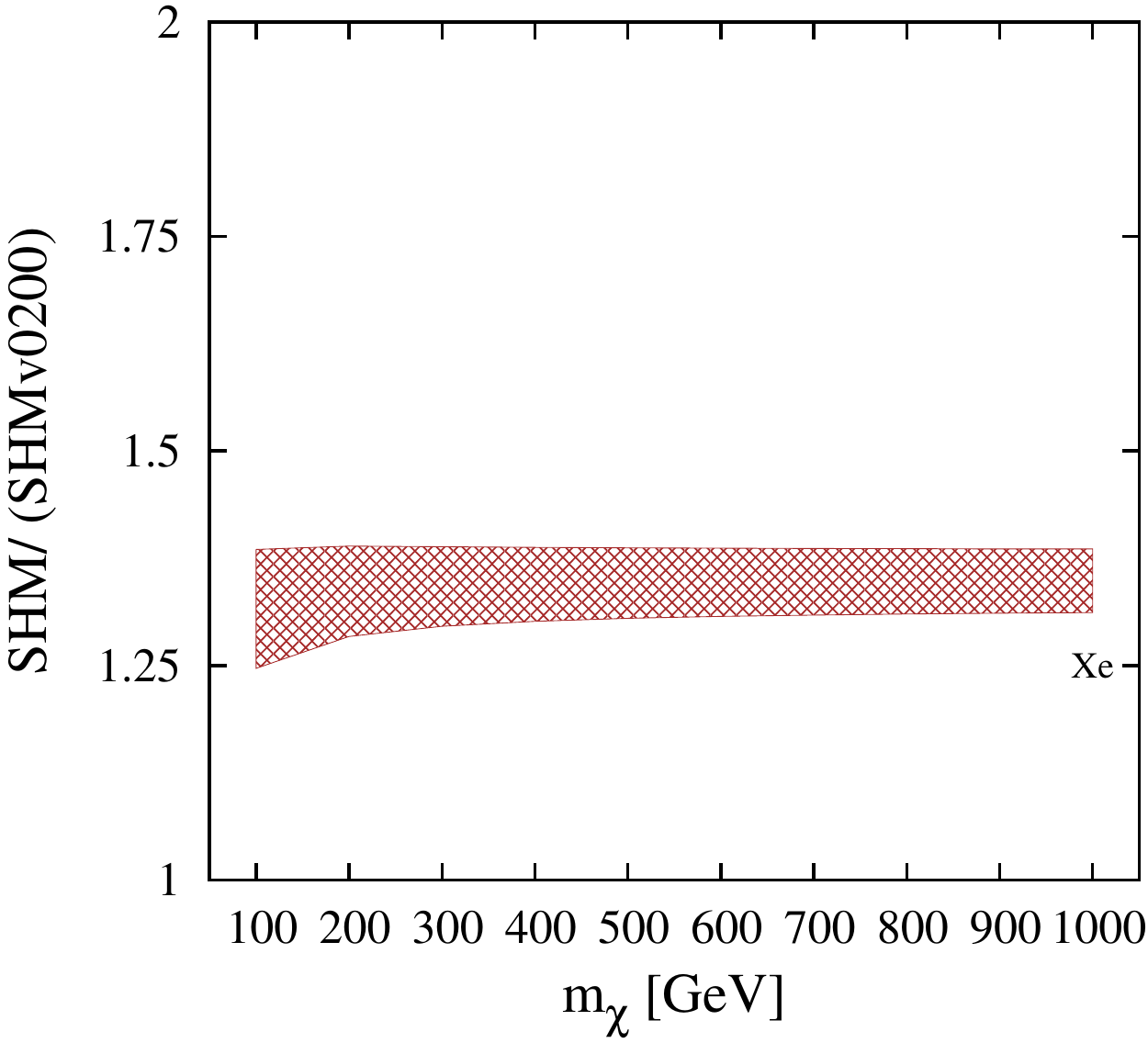}
\includegraphics[angle=0.0,width=0.48\columnwidth]{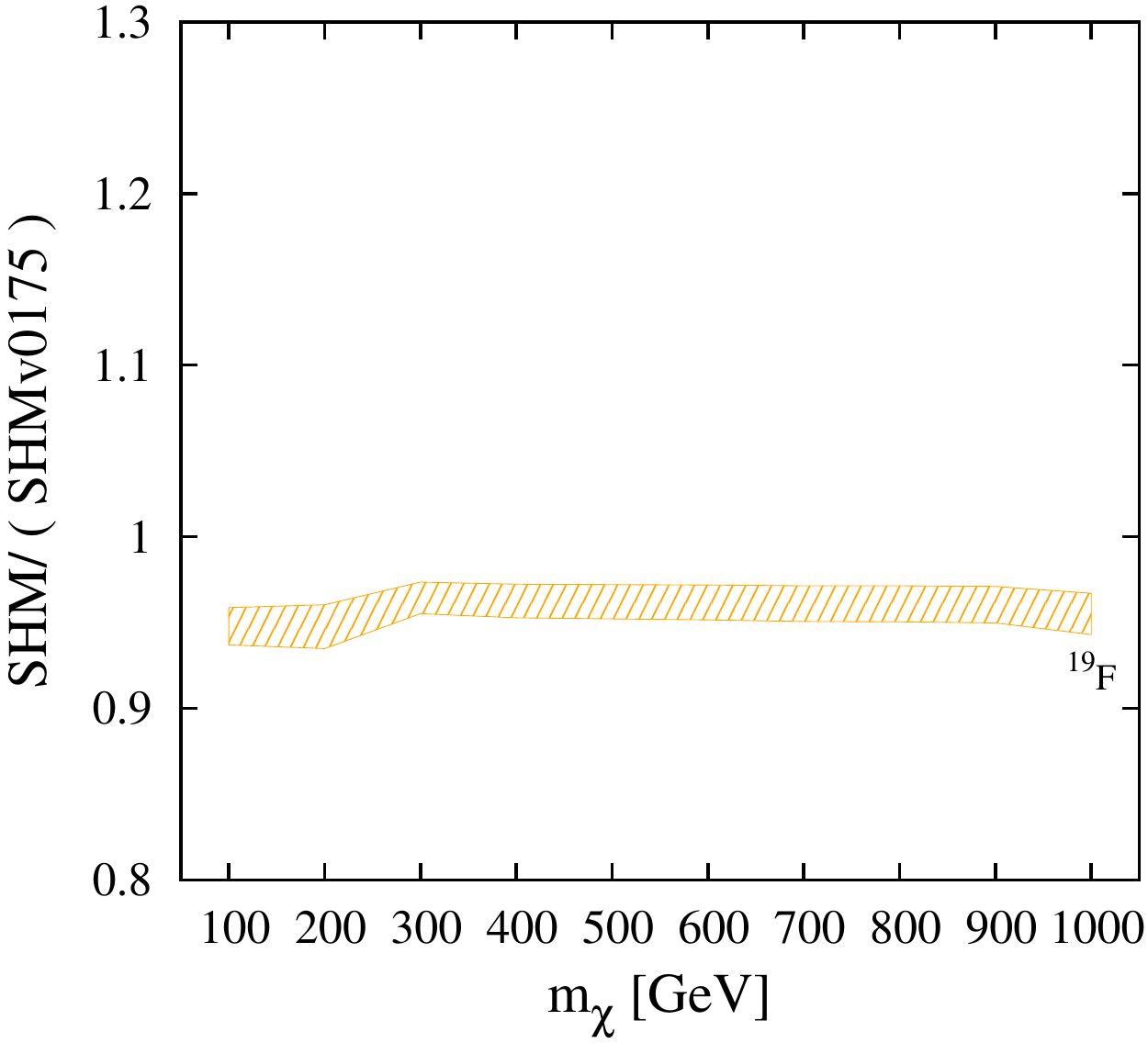}
\includegraphics[angle=0.0,width=0.48\columnwidth]{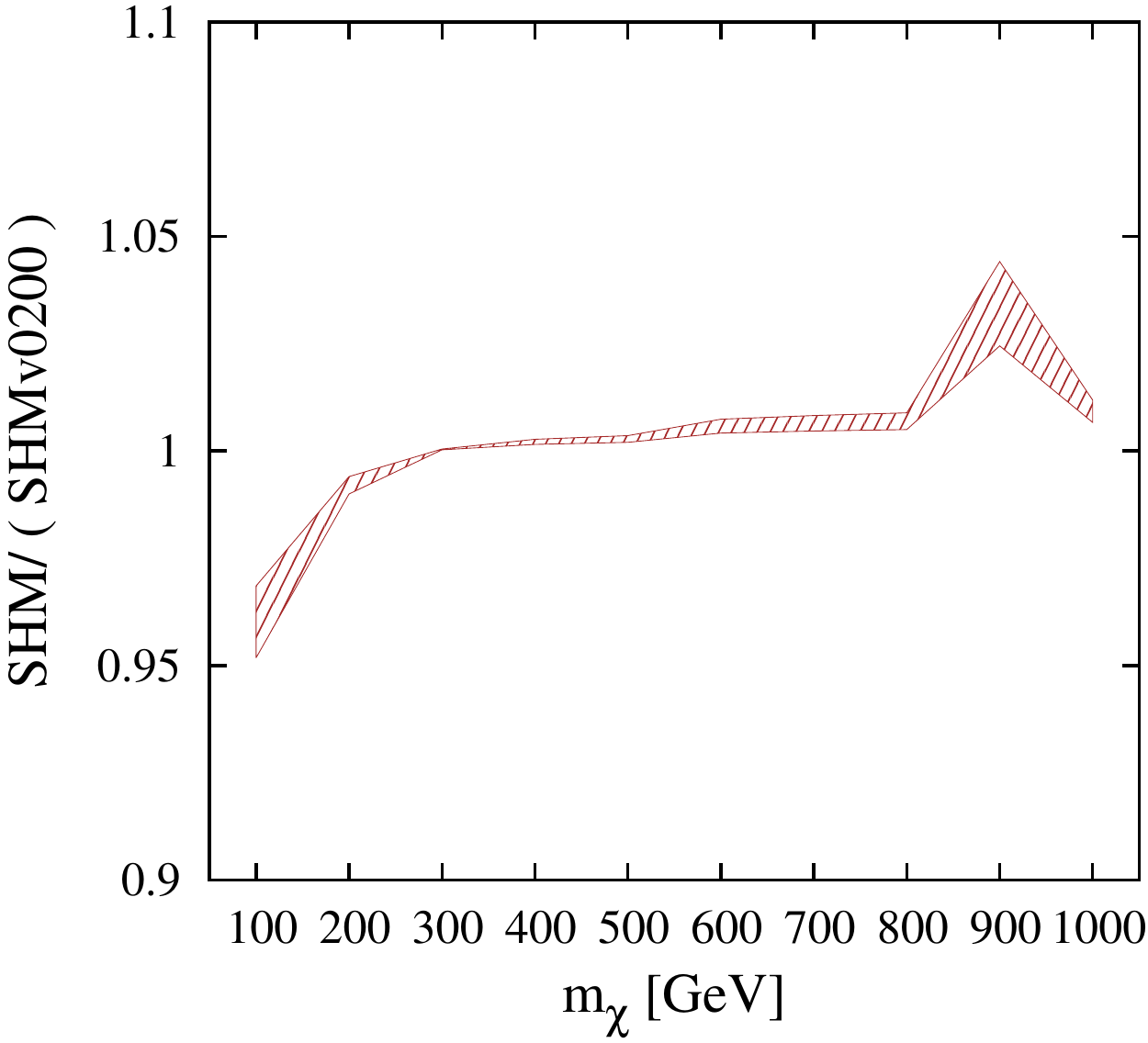}
\caption{Same as Fig.\,\ref{fig:v175 v200 19F} but for Xe.}
\label{fig:v175 v200 Xe}
\end{figure}

As explained earlier, it is possible to explain the results in this paper analytically by focusing on the ``effective $v_0$" of the dark matter velocity distribution.  To explain the trend with varying $v_0$, we will now show the results for dark matter velocity distribution with different $v_0$.  For example, the results obtained in this paper using the Mao et al., dark matter velocity distribution for the E9 halo can be well approximated by assuming an SHM velocity distribution with $v_0 = 150$ km s$^{-1}$.  For pedagogical purposes, we will now display our results for a dark matter velocity distribution following the SHM form with $v_0 = 175$ km s$^{-1}$ and $v_0 = 200$ km s$^{-1}$.  These are shown in Figs.\,\ref{fig:v175 v200 19F} and \ref{fig:v175 v200 Xe}.  These ratios show that both the forward - backward ratio and the ring is more pronounced for a dark matter velocity profile with a smaller ``effective $v_0$" as analytically explained earlier.  We want to remark that the two dark matter velocity distributions used in Figs.\,\ref{fig:v175 v200 19F} and \ref{fig:v175 v200 Xe} are not derived from hydrodynamical simulations, but are used for explanatory purposes.

\section{Conclusion}
\label{sec:conclusion}

Directional detection of dark matter is one of the most promising ways to unambiguously detect dark matter.  Although present constraints are weak, it is expected that near future technological progress will make them more competitive.  It has been pointed out that the forward-backward asymmetry and the ring-like structure of the maximum recoil rate can be used as an efficient discriminator between signal and background.  

Recent hydrodynamical simulations have shown that the dark matter velocity profile differs substantially from the Standard Halo Model.  We consider the SHM, the Mao et al. fit and the Maxwellian fit to halos E9 and E11 from the EAGLE HR and the APOSTLE IR simulations (see Fig.\,\ref{fig:Comparison velocity profile}).  The Mao et al. fit has a larger and smaller number of dark matter particles at low and high velocities respectively.

The effect of these different velocity profiles is shown in Fig.\,\ref{fig:differential recoil rate} for the $^{19}$F and Xe nuclei.  The Mao et al. profile shows a much more distinct forward-backward asymmetry and consequently the presence of a ring.  It is evident from this figure that fewer events will be required to estimate the forward-backward asymmetry and the presence of a ring when the Mao et al. dark matter velocity profile is considered.

The ratio of the number of events required for the discrimination of the forward-backward asymmetry and the ring is shown in Figs.\,\ref{fig:forward backward 19F}, \ref{fig:forward backward Xe}, \ref{fig:ring 19F} and \ref{fig:ring Xe}.  The number of events required for a 3$\sigma$ determination of the forward-backward asymmetry and for the evidence of a ring for the Mao et al. profile is $\sim$2-3 times less than that for SHM for both $^{19}$F and Xe target. 

It is important to use realistic dark matter velocity distributions while interpreting dark matter direct detection experiments.  In this work we studied the impact of these velocity distributions on directional dark matter experiments.  The realistic dark matter velocity distributions produce a much more dramatic  forward-backward asymmetry in these experiments.  The ring of the maximum dark matter recoil rate is also much more prominent if the Mao et al. velocity distribution is realized in nature.  Along with other detection modes of dark matter (see for e.g.,\,\cite{Dasgupta:2012bd,Laha:2012fg,Ng:2013xha,Murase:2015gea,Speckhard:2015eva,Chowdhury:2016bxs}), we hope that the use of these realistic dark matter velocity distributions will improve our understanding of the dark sector physics.

\section*{Acknowledgments} 
We thank Yao-Yuan Mao and Risa Wechsler for discussion.  R.L. is supported by DOE Contract DE-AC02-76SF00515.

\bibliographystyle{kp}
\interlinepenalty=10000
\tolerance=100
\bibliography{Bibliography/references}

\begingroup\raggedright\begin{thebibliography}{75}
\expandafter\ifx\csname natexlab\endcsname\relax\def\natexlab#1{#1}\fi

\bibitem[Ade et~al.(2015)]{Planck:2015xua}
{\bfseries Planck} Collaboration, P.~Ade {\em et~al.}, ``{Planck 2015 results.
  XIII. Cosmological parameters}'',
 \href{http://xxx.lanl.gov/abs/1502.01589}{ arXiv:1502.01589}.

\bibitem[Steigman(2007)]{Steigman:2007xt}
G.~Steigman, ``{Primordial Nucleosynthesis in the Precision Cosmology Era}'',
  {\em Ann.Rev.Nucl.Part.Sci.} {\bfseries 57} (2007) 463--491,
 \href{http://xxx.lanl.gov/abs/0712.1100}{ arXiv:0712.1100}.

\bibitem[Strigari et~al.(2014)Strigari, Frenk, and White]{Strigari:2014yea}
L.~E. Strigari, C.~S. Frenk, and S.~D.~M. White, ``{Dynamical models for the
  Sculptor dwarf spheroidal in a Lambda CDM universe}'',
 \href{http://xxx.lanl.gov/abs/1406.6079}{ arXiv:1406.6079}.

\bibitem[Bhattacharjee et~al.(2013)Bhattacharjee, Chaudhury, Kundu, and
  Majumdar]{Bhattacharjee:2012xm}
P.~Bhattacharjee, S.~Chaudhury, S.~Kundu, and S.~Majumdar, ``{Sizing-up the
  WIMPs of Milky Way : Deriving the velocity distribution of Galactic Dark
  Matter particles from the rotation curve data}'', {\em Phys.Rev.} {\bfseries
  D87} (2013) 083525,
 \href{http://xxx.lanl.gov/abs/1210.2328}{ arXiv:1210.2328}.

\bibitem[Marrodán~Undagoitia and Rauch(2016)]{Undagoitia:2015gya}
T.~Marrodán~Undagoitia and L.~Rauch, ``{Dark matter direct-detection
  experiments}'', {\em J. Phys.} {\bfseries G43} (2016), no.~1, 013001,
 \href{http://xxx.lanl.gov/abs/1509.08767}{ arXiv:1509.08767}.

\bibitem[Busoni et~al.(2016)]{Boveia:2016mrp}
G.~Busoni {\em et~al.}, ``{Recommendations on presenting LHC searches for
  missing transverse energy signals using simplified $s$-channel models of dark
  matter}'',
 \href{http://xxx.lanl.gov/abs/1603.04156}{ arXiv:1603.04156}.

\bibitem[Klasen et~al.(2015)Klasen, Pohl, and Sigl]{Klasen:2015uma}
M.~Klasen, M.~Pohl, and G.~Sigl, ``{Indirect and direct search for dark
  matter}'', {\em Prog. Part. Nucl. Phys.} {\bfseries 85} (2015) 1--32,
 \href{http://xxx.lanl.gov/abs/1507.03800}{ arXiv:1507.03800}.

\bibitem[Catena and Ullio(2012)]{Catena:2011kv}
R.~Catena and P.~Ullio, ``{The local dark matter phase-space density and impact
  on WIMP direct detection}'', {\em JCAP} {\bfseries 1205} (2012) 005,
 \href{http://xxx.lanl.gov/abs/1111.3556}{ arXiv:1111.3556}.

\bibitem[Herrero-Garcia(2015)]{Herrero-Garcia:2015kga}
J.~Herrero-Garcia, ``{Halo-independent tests of dark matter annual modulation
  signals}'', {\em JCAP} {\bfseries 1509} (2015), no.~09, 012,
 \href{http://xxx.lanl.gov/abs/1506.03503}{ arXiv:1506.03503}.

\bibitem[Del~Nobile et~al.(2015)Del~Nobile, Gelmini, Georgescu, and
  Huh]{DelNobile:2015lxa}
E.~Del~Nobile, G.~B. Gelmini, A.~Georgescu, and J.-H. Huh, ``{Reevaluation of
  spin-dependent WIMP-proton interactions as an explanation of the DAMA
  data}'', {\em JCAP} {\bfseries 1508} (2015), no.~08, 046,
 \href{http://xxx.lanl.gov/abs/1502.07682}{ arXiv:1502.07682}.

\bibitem[Catena et~al.(2016)Catena, Ibarra, and Wild]{Catena:2016hoj}
R.~Catena, A.~Ibarra, and S.~Wild, ``{DAMA confronts null searches in the
  effective theory of dark matter-nucleon interactions}'', {\em JCAP}
  {\bfseries 1605} (2016), no.~05, 039,
 \href{http://xxx.lanl.gov/abs/1602.04074}{ arXiv:1602.04074}.

\bibitem[Scopel and Yoon(2016)]{Scopel:2015eoh}
S.~Scopel and K.-H. Yoon, ``{Inelastic dark matter with spin-dependent
  couplings to protons and large modulation fractions in DAMA}'', {\em JCAP}
  {\bfseries 1602} (2016), no.~02, 050,
 \href{http://xxx.lanl.gov/abs/1512.00593}{ arXiv:1512.00593}.

\bibitem[Yang(2016)]{Yang:2016wrl}
K.-C. Yang, ``{Fermionic Dark Matter through a Light Pseudoscalar Portal: Hints
  from the DAMA Results}'', {\em Phys. Rev.} {\bfseries D94} (2016), no.~3,
  035028,
 \href{http://xxx.lanl.gov/abs/1604.04979}{ arXiv:1604.04979}.

\bibitem[Spergel(1988)]{Spergel:1987kx}
D.~N. Spergel, ``{The Motion of the Earth and the Detection of Wimps}'', {\em
  Phys.Rev.} {\bfseries D37} (1988)
1353.

\bibitem[Copi et~al.(1999)Copi, Heo, and Krauss]{Copi:1999pw}
C.~J. Copi, J.~Heo, and L.~M. Krauss, ``{Directional sensitivity, WIMP
  detection, and the galactic halo}'', {\em Phys.Lett.} {\bfseries B461} (1999)
  43--48,
 \href{http://xxx.lanl.gov/abs/hep-ph/9904499}{ arXiv:hep-ph/9904499}.

\bibitem[Copi and Krauss(2001)]{Copi:2000tv}
C.~J. Copi and L.~M. Krauss, ``{Angular signatures for galactic halo WIMP
  scattering in direct detectors: Prospects and challenges}'', {\em Phys.Rev.}
  {\bfseries D63} (2001) 043507,
 \href{http://xxx.lanl.gov/abs/astro-ph/0009467}{ arXiv:astro-ph/0009467}.

\bibitem[Gondolo(2002)]{Gondolo:2002np}
P.~Gondolo, ``{Recoil momentum spectrum in directional dark matter
  detectors}'', {\em Phys.Rev.} {\bfseries D66} (2002) 103513,
 \href{http://xxx.lanl.gov/abs/hep-ph/0209110}{ arXiv:hep-ph/0209110}.

\bibitem[Morgan et~al.(2005)Morgan, Green, and Spooner]{Morgan:2004ys}
B.~Morgan, A.~M. Green, and N.~J. Spooner, ``{Directional statistics for WIMP
  direct detection}'', {\em Phys.Rev.} {\bfseries D71} (2005) 103507,
 \href{http://xxx.lanl.gov/abs/astro-ph/0408047}{ arXiv:astro-ph/0408047}.

\bibitem[Billard et~al.(2010)Billard, Mayet, Macias-Perez, and
  Santos]{Billard:2009mf}
J.~Billard, F.~Mayet, J.~Macias-Perez, and D.~Santos, ``{Directional detection
  as a strategy to discover galactic Dark Matter}'', {\em Phys.Lett.}
  {\bfseries B691} (2010) 156--162,
 \href{http://xxx.lanl.gov/abs/0911.4086}{ arXiv:0911.4086}.

\bibitem[Ahlen et~al.(2010)Ahlen, Afshordi, Battat, Billard, Bozorgnia,
  et~al.]{Ahlen:2009ev}
S.~Ahlen, N.~Afshordi, J.~Battat, J.~Billard, N.~Bozorgnia, {\em et~al.},
  ``{The case for a directional dark matter detector and the status of current
  experimental efforts}'', {\em Int.J.Mod.Phys.} {\bfseries A25} (2010) 1--51,
 \href{http://xxx.lanl.gov/abs/0911.0323}{ arXiv:0911.0323}.

\bibitem[Green and Morgan(2010)]{Green:2010zm}
A.~M. Green and B.~Morgan, ``{The median recoil direction as a WIMP directional
  detection signal}'', {\em Phys.Rev.} {\bfseries D81} (2010) 061301,
 \href{http://xxx.lanl.gov/abs/1002.2717}{ arXiv:1002.2717}.

\bibitem[Lee and Peter(2012)]{Lee:2012pf}
S.~K. Lee and A.~H. Peter, ``{Probing the Local Velocity Distribution of WIMP
  Dark Matter with Directional Detectors}'', {\em JCAP} {\bfseries 1204} (2012)
  029,
 \href{http://xxx.lanl.gov/abs/1202.5035}{ arXiv:1202.5035}.

\bibitem[Grothaus et~al.(2014)Grothaus, Fairbairn, and
  Monroe]{Grothaus:2014hja}
P.~Grothaus, M.~Fairbairn, and J.~Monroe, ``{Directional Dark Matter Detection
  Beyond the Neutrino Bound}'', {\em Phys.Rev.} {\bfseries D90} (2014), no.~5,
  055018,
 \href{http://xxx.lanl.gov/abs/1406.5047}{ arXiv:1406.5047}.

\bibitem[O'Hare and Green(2014)]{O'Hare:2014oxa}
C.~A.~J. O'Hare and A.~M. Green, ``{Directional detection of dark matter
  streams}'', {\em Phys.Rev.} {\bfseries D90} (2014), no.~12, 123511,
 \href{http://xxx.lanl.gov/abs/1410.2749}{ arXiv:1410.2749}.

\bibitem[O'Hare et~al.(2015)O'Hare, Green, Billard, Figueroa-Feliciano, and
  Strigari]{O'Hare:2015mda}
C.~A.~J. O'Hare, A.~M. Green, J.~Billard, E.~Figueroa-Feliciano, and L.~E.
  Strigari, ``{Readout strategies for directional dark matter detection beyond
  the neutrino background}'', {\em Phys. Rev.} {\bfseries D92} (2015), no.~6,
  063518,
 \href{http://xxx.lanl.gov/abs/1505.08061}{ arXiv:1505.08061}.

\bibitem[Laha(2015)]{Laha:2015yoa}
R.~Laha, ``{Directional detection of dark matter in universal bound states}'',
  {\em Phys. Rev.} {\bfseries D92} (2015) 083509,
 \href{http://xxx.lanl.gov/abs/1505.02772}{ arXiv:1505.02772}.

\bibitem[Mayet et~al.(2016)]{Mayet:2016zxu}
F.~Mayet {\em et~al.}, ``{A review of the discovery reach of directional Dark
  Matter detection}'', {\em Phys. Rept.} {\bfseries 627} (2016) 1--49,
 \href{http://xxx.lanl.gov/abs/1602.03781}{ arXiv:1602.03781}.

\bibitem[Kavanagh(2015)]{Kavanagh:2015aqa}
B.~J. Kavanagh, ``{Discretising the velocity distribution for directional dark
  matter experiments}'',
 \href{http://xxx.lanl.gov/abs/1502.04224}{ arXiv:1502.04224}.

\bibitem[Kavanagh and O'Hare(2016)]{Kavanagh:2016xfi}
B.~J. Kavanagh and C.~A.~J. O'Hare, ``{Reconstructing the three-dimensional
  local dark matter velocity distribution}'',
 \href{http://xxx.lanl.gov/abs/1609.08630}{ arXiv:1609.08630}.

\bibitem[Daw et~al.(2012)]{Daw:2011wq}
E.~Daw {\em et~al.}, ``{The DRIFT Directional Dark Matter Experiments}'', {\em
  EAS Publ. Ser.} {\bfseries 53} (2012) 11--18,
 \href{http://xxx.lanl.gov/abs/1110.0222}{ arXiv:1110.0222}.

\bibitem[Battat et~al.(2014)]{Battat:2014van}
{\bfseries DRIFT Collaboration} Collaboration, J.~Battat {\em et~al.}, ``{First
  background-free limit from a directional dark matter experiment: results from
  a fully fiducialised DRIFT detector}'',
 \href{http://xxx.lanl.gov/abs/1410.7821}{ arXiv:1410.7821}.

\bibitem[Jaegle et~al.(2012)Jaegle, Feng, Ross, Yamaoka, Vahsen, Feng, Ross,
  Yamaoka, and Vahsen]{Jaegle:2011rn}
I.~Jaegle, H.~Feng, S.~Ross, J.~Yamaoka, S.~E. Vahsen, H.~Feng, S.~Ross,
  J.~Yamaoka, and S.~E. Vahsen, ``{Simulation of the Directional Dark Matter
  Detector ($D^{3}$) and Directional Neutron Observer (DiNO)}'', {\em EAS Publ.
  Ser.} {\bfseries 53} (2012) 111--118,
 \href{http://xxx.lanl.gov/abs/1110.3444}{ arXiv:1110.3444}.

\bibitem[Vahsen et~al.(2012)Vahsen, Feng, Garcia-Sciveres, Jaegle, Kadyk,
  et~al.]{Vahsen:2011qx}
S.~Vahsen, H.~Feng, M.~Garcia-Sciveres, I.~Jaegle, J.~Kadyk, {\em et~al.},
  ``{The Directional Dark Matter Detector ($D^{3}$)}'', {\em EAS Publ.Ser.}
  {\bfseries 53} (2012) 43--50,
 \href{http://xxx.lanl.gov/abs/1110.3401}{ arXiv:1110.3401}.

\bibitem[Monroe(2012{\natexlab{a}})]{Monroe:2012qma}
{\bfseries DMTPC Collaboration} Collaboration, J.~Monroe, ``{Status and
  Prospects of the DMTPC Directional Dark Matter Experiment}'', {\em EAS
  Publ.Ser.} {\bfseries 53} (2012){\natexlab{a}}
19--24.

\bibitem[Monroe(2012{\natexlab{b}})]{Monroe:2011er}
{\bfseries DMTPC} Collaboration, J.~Monroe, ``{Status and Prospects of the
  DMTPC Directional Dark Matter Experiment}'', {\em AIP Conf. Proc.} {\bfseries
  1441} (2012){\natexlab{b}} 515--517,
 \href{http://xxx.lanl.gov/abs/1111.0220}{ arXiv:1111.0220}.

\bibitem[Leyton(2016)]{Leyton:2016nit}
{\bfseries DMTPC} Collaboration, M.~Leyton, ``{Directional dark matter
  detection with the DMTPC m$^3$ experiment}'', {\em J. Phys. Conf. Ser.}
  {\bfseries 718} (2016), no.~4,
042035.

\bibitem[Miuchi et~al.(2010)]{Miuchi:2010hn}
K.~Miuchi {\em et~al.}, ``{First underground results with NEWAGE-0.3a
  direction-sensitive dark matter detector}'', {\em Phys. Lett.} {\bfseries
  B686} (2010) 11--17,
 \href{http://xxx.lanl.gov/abs/1002.1794}{ arXiv:1002.1794}.

\bibitem[Nakamura et~al.(2012)Nakamura, Miuchi, Iwaki, Kubo, Mizumoto,
  et~al.]{Miuchi:2011qw}
K.~Nakamura, K.~Miuchi, S.~Iwaki, H.~Kubo, T.~Mizumoto, {\em et~al.},
  ``{NEWAGE}'', {\em J.Phys.Conf.Ser.} {\bfseries 375} (2012) 012013,
 \href{http://xxx.lanl.gov/abs/1109.3099}{ arXiv:1109.3099}.

\bibitem[Riffard et~al.(2013)Riffard, Billard, Bosson, Bourrion, Guillaudin,
  et~al.]{Riffard:2013psa}
Q.~Riffard, J.~Billard, G.~Bosson, O.~Bourrion, O.~Guillaudin, {\em et~al.},
  ``{Dark Matter directional detection with MIMAC}'',
 \href{http://xxx.lanl.gov/abs/1306.4173}{ arXiv:1306.4173}.

\bibitem[Riffard et~al.(2016)]{Riffard:2016mgw}
Q.~Riffard {\em et~al.}, ``{MIMAC low energy electron-recoil discrimination
  measured with fast neutrons}'', {\em JINST} {\bfseries 11} (2016), no.~08,
  P08011,
 \href{http://xxx.lanl.gov/abs/1602.01738}{ arXiv:1602.01738}.

\bibitem[Naka and Miuchi(2013)]{Naka:2013nla}
T.~Naka and K.~Miuchi, eds., {\em {Proceedings, 4th Workshop on Directional
  Detection of Dark Matter (CYGNUS 2013)}}, vol.~469.
\newblock
2013.
\newblock

\bibitem[Cappella et~al.(2013)]{Cappella:2013rua}
F.~Cappella {\em et~al.}, ``{On the potentiality of the $ZnWO_{4}$ anisotropic
  detectors to measure the directionality of Dark Matter}'', {\em Eur. Phys.
  J.} {\bfseries C73} (2013), no.~1,
2276.

\bibitem[Capparelli et~al.(2015)Capparelli, Cavoto, Mazzilli, and
  Polosa]{Capparelli:2014lua}
L.~M. Capparelli, G.~Cavoto, D.~Mazzilli, and A.~D. Polosa, ``{Directional Dark
  Matter Searches with Carbon Nanotubes}'', {\em Phys. Dark Univ.} {\bfseries
  9-10} (2015) 24--30,  \href{http://xxx.lanl.gov/abs/1412.8213}{
  arXiv:1412.8213},
[Erratum: Phys. Dark Univ.11,79(2016)].

\bibitem[Aleksandrov et~al.(2016)]{Aleksandrov:2016fyr}
{\bfseries NEWS} Collaboration, A.~Aleksandrov {\em et~al.}, ``{NEWS: Nuclear
  Emulsions for WIMP Search}'',
 \href{http://xxx.lanl.gov/abs/1604.04199}{ arXiv:1604.04199}.

\bibitem[Nygren(2013)]{Nygren:2013nda}
D.~Nygren, ``{Columnar recombination: a tool for nuclear recoil directional
  sensitivity in a xenon-based direct detection WIMP search}'', {\em
  J.Phys.Conf.Ser.} {\bfseries 460} (2013)
012006.

\bibitem[Gehman et~al.(2013)Gehman, Goldschmidt, Nygren, Oliveira, and
  Renner]{Gehman:2013mra}
V.~Gehman, A.~Goldschmidt, D.~Nygren, C.~Oliveira, and J.~Renner, ``{A plan for
  directional dark matter sensitivity in high-pressure xenon detectors through
  the addition of wavelength shifting gaseous molecules}'', {\em JINST}
  {\bfseries 8} (2013)
C10001.

\bibitem[Mohlabeng et~al.(2015)Mohlabeng, Kong, Li, Para, and
  Yoo]{Mohlabeng:2015efa}
G.~Mohlabeng, K.~Kong, J.~Li, A.~Para, and J.~Yoo, ``{Dark Matter
  Directionality Revisited with a High Pressure Xenon Gas Detector}'',
 \href{http://xxx.lanl.gov/abs/1503.03937}{ arXiv:1503.03937}.

\bibitem[Li(2015)]{Li:2015zga}
J.~Li, ``{Directional dark matter by polar angle direct detection and
  application of columnar recombination}'',
 \href{http://xxx.lanl.gov/abs/1503.07320}{ arXiv:1503.07320}.

\bibitem[Bozorgnia et~al.(2012)Bozorgnia, Gelmini, and
  Gondolo]{Bozorgnia:2011vc}
N.~Bozorgnia, G.~B. Gelmini, and P.~Gondolo, ``{Ring-like features in
  directional dark matter detection}'', {\em JCAP} {\bfseries 1206} (2012) 037,
 \href{http://xxx.lanl.gov/abs/1111.6361}{ arXiv:1111.6361}.

\bibitem[Laha and Braaten(2014)]{Laha:2013gva}
R.~Laha and E.~Braaten, ``{Direct detection of dark matter in universal bound
  states}'', {\em Phys.Rev.} {\bfseries D89} (2014), no.~10, 103510,
 \href{http://xxx.lanl.gov/abs/1311.6386}{ arXiv:1311.6386}.

\bibitem[Ling et~al.(2010)Ling, Nezri, Athanassoula, and Teyssier]{Ling:2009eh}
F.~S. Ling, E.~Nezri, E.~Athanassoula, and R.~Teyssier, ``{Dark Matter Direct
  Detection Signals inferred from a Cosmological N-body Simulation with
  Baryons}'', {\em JCAP} {\bfseries 1002} (2010) 012,
 \href{http://xxx.lanl.gov/abs/0909.2028}{ arXiv:0909.2028}.

\bibitem[Kuhlen et~al.(2014)Kuhlen, Pillepich, Guedes, and
  Madau]{Kuhlen:2013tra}
M.~Kuhlen, A.~Pillepich, J.~Guedes, and P.~Madau, ``{The Distribution of Dark
  Matter in the Milky Way's Disk}'', {\em Astrophys. J.} {\bfseries 784} (2014)
  161,
 \href{http://xxx.lanl.gov/abs/1308.1703}{ arXiv:1308.1703}.

\bibitem[Butsky et~al.(2015)Butsky, Maccio, Dutton, Wang, Stinson, Penzo, Kang,
  Keller, and Wadsley]{Butsky:2015pya}
I.~Butsky, A.~V. Maccio, A.~A. Dutton, L.~Wang, G.~S. Stinson, C.~Penzo,
  X.~Kang, B.~W. Keller, and J.~Wadsley, ``{NIHAO project II: Halo shape,
  phase-space density and velocity distribution of dark matter in galaxy
  formation simulations}'',
 \href{http://xxx.lanl.gov/abs/1503.04814}{ arXiv:1503.04814}.

\bibitem[Bozorgnia et~al.(2016)Bozorgnia, Calore, Schaller, Lovell, Bertone,
  Frenk, Crain, Navarro, Schaye, and Theuns]{Bozorgnia:2016ogo}
N.~Bozorgnia, F.~Calore, M.~Schaller, M.~Lovell, G.~Bertone, C.~S. Frenk, R.~A.
  Crain, J.~F. Navarro, J.~Schaye, and T.~Theuns, ``{Simulated Milky Way
  analogues: implications for dark matter direct searches}'', {\em JCAP}
  {\bfseries 1605} (2016), no.~05, 024,
 \href{http://xxx.lanl.gov/abs/1601.04707}{ arXiv:1601.04707}.

\bibitem[Kelso et~al.(2016)Kelso, Savage, Valluri, Freese, Stinson, and
  Bailin]{Kelso:2016qqj}
C.~Kelso, C.~Savage, M.~Valluri, K.~Freese, G.~S. Stinson, and J.~Bailin,
  ``{The impact of baryons on the direct detection of dark matter}'', {\em
  JCAP} {\bfseries 1608} (2016) 071,
 \href{http://xxx.lanl.gov/abs/1601.04725}{ arXiv:1601.04725}.

\bibitem[Sloane et~al.(2016)Sloane, Buckley, Brooks, and
  Governato]{Sloane:2016kyi}
J.~D. Sloane, M.~R. Buckley, A.~M. Brooks, and F.~Governato, ``{Assessing
  Astrophysical Uncertainties in Direct Detection with Galaxy Simulations}'',
 \href{http://xxx.lanl.gov/abs/1601.05402}{ arXiv:1601.05402}.

\bibitem[Petersen et~al.(2016)Petersen, Katz, and Weinberg]{Petersen:2016vck}
M.~S. Petersen, N.~Katz, and M.~D. Weinberg, ``{The Dynamical Response of Dark
  Matter to Galaxy Evolution Affects Direct-Detection Experiments}'',
 \href{http://xxx.lanl.gov/abs/1609.01307}{ arXiv:1609.01307}.

\bibitem[Piffl et~al.(2014)]{Piffl:2013mla}
T.~Piffl {\em et~al.}, ``{The RAVE survey: the Galactic escape speed and the
  mass of the Milky Way}'', {\em Astron. Astrophys.} {\bfseries 562} (2014)
  A91,
 \href{http://xxx.lanl.gov/abs/1309.4293}{ arXiv:1309.4293}.

\bibitem[Schaye et~al.(2015)]{Schaye:2014tpa}
J.~Schaye {\em et~al.}, ``{The EAGLE project: Simulating the evolution and
  assembly of galaxies and their environments}'', {\em Mon. Not. Roy. Astron.
  Soc.} {\bfseries 446} (2015) 521--554,
 \href{http://xxx.lanl.gov/abs/1407.7040}{ arXiv:1407.7040}.

\bibitem[Crain et~al.(2015)]{Crain:2015poa}
R.~A. Crain {\em et~al.}, ``{The EAGLE simulations of galaxy formation:
  calibration of subgrid physics and model variations}'', {\em Mon. Not. Roy.
  Astron. Soc.} {\bfseries 450} (2015), no.~2, 1937--1961,
 \href{http://xxx.lanl.gov/abs/1501.01311}{ arXiv:1501.01311}.

\bibitem[Sawala et~al.(2016)]{Sawala:2015cdf}
T.~Sawala {\em et~al.}, ``{The APOSTLE simulations: solutions to the Local
  Group's cosmic puzzles}'', {\em Mon. Not. Roy. Astron. Soc.} {\bfseries 457}
  (2016), no.~2, 1931--1943,
 \href{http://xxx.lanl.gov/abs/1511.01098}{ arXiv:1511.01098}.

\bibitem[{Fattahi} et~al.(2016){Fattahi}, {Navarro}, {Sawala}, {Frenk}, {Oman},
  {Crain}, {Furlong}, {Schaller}, {Schaye}, {Theuns}, and
  {Jenkins}]{2016MNRAS.457..844F}
A.~{Fattahi}, J.~F. {Navarro}, T.~{Sawala}, C.~S. {Frenk}, K.~A. {Oman}, R.~A.
  {Crain}, M.~{Furlong}, M.~{Schaller}, J.~{Schaye}, T.~{Theuns}, and
  A.~{Jenkins}, ``{The APOSTLE project: Local Group kinematic mass constraints
  and simulation candidate selection}'', {\em Monthly Notices of the Royal
  Astronomical Society} {\bfseries 457} (2016) 844--856,
  \href{http://xxx.lanl.gov/abs/1507.03643}{ arXiv:1507.03643}.

\bibitem[Mao et~al.(2013)Mao, Strigari, Wechsler, Wu, and Hahn]{Mao:2012hf}
Y.-Y. Mao, L.~E. Strigari, R.~H. Wechsler, H.-Y. Wu, and O.~Hahn,
  ``{Halo-to-Halo Similarity and Scatter in the Velocity Distribution of Dark
  Matter}'', {\em Astrophys. J.} {\bfseries 764} (2013) 35,
 \href{http://xxx.lanl.gov/abs/1210.2721}{ arXiv:1210.2721}.

\bibitem[Lisanti et~al.(2011)Lisanti, Strigari, Wacker, and
  Wechsler]{Lisanti:2010qx}
M.~Lisanti, L.~E. Strigari, J.~G. Wacker, and R.~H. Wechsler, ``{The Dark
  Matter at the End of the Galaxy}'', {\em Phys. Rev.} {\bfseries D83} (2011)
  023519,
 \href{http://xxx.lanl.gov/abs/1010.4300}{ arXiv:1010.4300}.

\bibitem[Calore et~al.(2015)Calore, Bozorgnia, Lovell, Bertone, Schaller,
  Frenk, Crain, Schaye, Theuns, and Trayford]{Calore:2015oya}
F.~Calore, N.~Bozorgnia, M.~Lovell, G.~Bertone, M.~Schaller, C.~S. Frenk, R.~A.
  Crain, J.~Schaye, T.~Theuns, and J.~W. Trayford, ``{Simulated Milky Way
  analogues: implications for dark matter indirect searches}'', {\em JCAP}
  {\bfseries 1512} (2015), no.~12, 053,
 \href{http://xxx.lanl.gov/abs/1509.02164}{ arXiv:1509.02164}.

\bibitem[Catena(2015)]{Catena:2015vpa}
R.~Catena, ``{Dark matter directional detection in non-relativistic effective
  theories}'', {\em JCAP} {\bfseries 1507} (2015), no.~07, 026,
 \href{http://xxx.lanl.gov/abs/1505.06441}{ arXiv:1505.06441}.

\bibitem[Kavanagh(2015)]{Kavanagh:2015jma}
B.~J. Kavanagh, ``{New directional signatures from the nonrelativistic
  effective field theory of dark matter}'', {\em Phys. Rev.} {\bfseries D92}
  (2015), no.~2, 023513,
 \href{http://xxx.lanl.gov/abs/1505.07406}{ arXiv:1505.07406}.

\bibitem[Battat et~al.(2016)]{Battat:2016pap}
J.~B.~R. Battat {\em et~al.}, ``{Readout technologies for directional WIMP Dark
  Matter detection}'', {\em Phys. Rept.} {\bfseries 662} (2016) 1--46,
 \href{http://xxx.lanl.gov/abs/1610.02396}{ arXiv:1610.02396}.

\bibitem[Bednyakov and Simkovic(2006)]{Bednyakov:2006ux}
V.~A. Bednyakov and F.~Simkovic, ``{Nuclear spin structure in dark matter
  search: The Finite momentum transfer limit}'', {\em Phys. Part. Nucl.}
  {\bfseries 37} (2006) S106--S128,
 \href{http://xxx.lanl.gov/abs/hep-ph/0608097}{ arXiv:hep-ph/0608097}.

\bibitem[Dasgupta and Laha(2012)]{Dasgupta:2012bd}
B.~Dasgupta and R.~Laha, ``{Neutrinos in IceCube/KM3NeT as probes of Dark
  Matter Substructures in Galaxy Clusters}'', {\em Phys. Rev.} {\bfseries D86}
  (2012) 093001,
 \href{http://xxx.lanl.gov/abs/1206.1322}{ arXiv:1206.1322}.

\bibitem[Laha et~al.(2013)Laha, Ng, Dasgupta, and Horiuchi]{Laha:2012fg}
R.~Laha, K.~C.~Y. Ng, B.~Dasgupta, and S.~Horiuchi, ``{Galactic center radio
  constraints on gamma-ray lines from dark matter annihilation}'', {\em Phys.
  Rev.} {\bfseries D87} (2013), no.~4, 043516,
 \href{http://xxx.lanl.gov/abs/1208.5488}{ arXiv:1208.5488}.

\bibitem[Ng et~al.(2014)Ng, Laha, Campbell, Horiuchi, Dasgupta, Murase, and
  Beacom]{Ng:2013xha}
K.~C.~Y. Ng, R.~Laha, S.~Campbell, S.~Horiuchi, B.~Dasgupta, K.~Murase, and
  J.~F. Beacom, ``{Resolving small-scale dark matter structures using
  multisource indirect detection}'', {\em Phys. Rev.} {\bfseries D89} (2014),
  no.~8, 083001,
 \href{http://xxx.lanl.gov/abs/1310.1915}{ arXiv:1310.1915}.

\bibitem[Murase et~al.(2015)Murase, Laha, Ando, and Ahlers]{Murase:2015gea}
K.~Murase, R.~Laha, S.~Ando, and M.~Ahlers, ``{Testing the Dark Matter Scenario
  for PeV Neutrinos Observed in IceCube}'', {\em Phys. Rev. Lett.} {\bfseries
  115} (2015), no.~7, 071301,
 \href{http://xxx.lanl.gov/abs/1503.04663}{ arXiv:1503.04663}.

\bibitem[Speckhard et~al.(2016)Speckhard, Ng, Beacom, and
  Laha]{Speckhard:2015eva}
E.~G. Speckhard, K.~C.~Y. Ng, J.~F. Beacom, and R.~Laha, ``{Dark Matter
  Velocity Spectroscopy}'', {\em Phys. Rev. Lett.} {\bfseries 116} (2016),
  no.~3, 031301,
 \href{http://xxx.lanl.gov/abs/1507.04744}{ arXiv:1507.04744}.

\bibitem[Chowdhury et~al.(2016)Chowdhury, Iyer, and Laha]{Chowdhury:2016bxs}
D.~Chowdhury, A.~M. Iyer, and R.~Laha, ``{Constraints on dark matter
  annihilation to fermions and a photon}'',
 \href{http://xxx.lanl.gov/abs/1601.06140}{ arXiv:1601.06140}.

\end{thebibliography}\endgroup

\end{document}